\documentclass[aps,preprint]{revtex4}%
\usepackage{amssymb}
\usepackage{amsfonts}
\usepackage{amsmath}
\usepackage{epstopdf}
\usepackage{natbib}
\usepackage{graphicx}%
\usepackage{float}
\restylefloat{table}
\providecommand{\U}[1]{\protect\rule{.1in}{.1in}}
\begin{document}
\preprint{ }
\title{Effective Field Theory for Bound State Reflection}
\author{Michelle Pine}
\affiliation{North Carolina State University}
\author{Dean Lee}
\affiliation{North Carolina State University}

\begin{abstract}
Elastic quantum bound-state reflection from a hard-wall boundary provides
direct information regarding the structure and compressibility of quantum
bound states. We discuss elastic quantum bound-state reflection and derive a
general theory for elastic reflection of shallow dimers from hard-wall
surfaces using effective field theory. We show that there is a small expansion
parameter for analytic calculations of the reflection scattering length. We
present a calculation up to second order in the effective Hamiltonian in one,
two, and three dimensions. We also provide numerical lattice results for all
three cases as a comparison with our effective field theory results. Finally,
we provide an analysis of the compressibility of the alpha particle confined to
a cubic lattice with vanishing Dirichlet boundaries.

\end{abstract}
\maketitle
\tableofcontents

\section{Introduction}

The study of confined quantum bound states has interesting applications in
many areas including nuclear structure calculations, experimental ultracold
atomic physics, and quantum dots and wells. Elastic reflection off of a hard
wall allows one to gain insight into the structure of a bound state as well as
how it behaves when confined or compressed. In the case of quantum dots and
wells, the bound state is comprised of an electron and hole confined inside a
two- or three- dimensional nanoscale semiconductor structure. Varying the
geometry of the structure allows increased control of current tunneling as
well as photon absorption and emission. See for example
Ref.~\cite{Takagahara:1992a,Marzin:1994a,Pfalz:2005a,He:2005a}. The theory
developed in this paper gives a universal result for the energy of Wannier
excitons in direct-band gap semiconductors as a function of the binding
energy, effective masses, and the geometry of the nanostructure.

In the case of ultra-cold atomic experiments,\ a quantum well is produced by
using lasers to create repulsive boundaries, thereby confining the atoms on an
optical lattice. Similar ideas have been proposed for quantum billiards
systems \cite{Montangero:2009}. Since they add no dimensionful scale to the
problem, hard-wall boundaries provide a probe of universal physics at large
scattering lengths. For example the scattering length for particle-particle
scattering is directly proportional to the scattering length for dimer-wall
reflection.\ In this paper we calculate this universal proportionality
constant in one, two and three spatial dimensions.

It is not possible, experimentally, to construct a hard-wall boundary for
protons and neutrons, so it may seem that the current discussion has no direct
connection with nuclear physics. A similar critique could be made of
L\"{u}scher's analysis of periodic boundaries in finite cubic volumes
\cite{Luscher:1985dn,Luscher:1986pf}. However, L\"{u}scher's analysis now
provides the theoretical framework for numerous calculations in lattice
quantum chromodynamics \cite{Beane:2005rj,Beane:2006mx,Bernard:2008ax}. While
it is not possible to construct a hard-wall boundary for protons and neutrons
in nuclear physics experiments, it is possible to do so in \textit{ab initio} numerical
calculations. In some cases such calculations can be used to verify phenomena
observed in experiments, and in other cases the calculations probe new physics
inaccessible in the laboratory. As with temperature and chemical potential,
boundaries can provide tunable control parameters for use in such
calculations. Hard-wall boundaries can and have been incorporated into
\textit{ab initio} numerical calculations
\cite{Epelbaum:2009pd,Epelbaum:2010xt}. Hard-wall boundaries are currently
being used to perform nuclear lattice calculations that probe properties such
as structure and elastic deformation of nuclei.\ The method is a useful
compliment to Luscher's periodic boundary analysis for bound states and
scattering states. By varying the boundary conditions one can calculate
quantities such as nuclear radii and quantities analogous to bulk and shear
modulus, which can then be compared with bulk modulus estimates inferred from
observed monopole resonance energies.

Recently there has been a great deal of interest in alpha-particle clusters
confined inside nuclei such as carbon-12
\cite{Tohsaki:2001an,Chernykh:2007a,Suzuki:2007wi}. Very recently \textit{ab
initio} lattice effective field theory calculations of carbon-12 have given
the\ energies of the ground state, the excited spin-2 state, and, for the
first time, the Hoyle state responsible for the formation of carbon in stellar
environments \cite{Epelbaum:2011md}.\ Furthermore, these calculations indicate
the presence of compressed, correlated alpha clusters. Since there are no
low-energy resonances of the alpha particle and little experimental data on
the compression of alpha particles, a better understanding of this phenomenon
would prove valuable. In this paper we analyze the compressibility of the
alpha particle.

In this paper we discuss the elastic reflection of quantum bound states off of
hard-wall boundaries. We begin with a short presentation of the general case,
where the number of dimensions and number of constituent particles is
completely arbitrary. We then develop a general theory from the principles of
effective field theory for the specific case of shallow two-body quantum bound
state reflection in one, two, and three dimensions.  Our main result is a 
derivation of the phase shift due to the scattering of a two-particle bound state on a hard \
wall in the adiabatic limit and up to second order in an expansion of the effective Hamiltonian. 
The effective field theory
results are presented for the cases of one, two, and three dimensions for
arbitrary mass ratio $m_{2}/m_{1}$ and compared to numerical results. Finally,
we provide an analysis of the compressibility of the alpha particle, summarize,
and consider further work to be done. A partial summary of our results was presented in a previous letter publication \cite{Lee:2010km}. Here we present the full details of our calculations. We note that very recently there has been work on two body systems in a finite volume with periodic boundaries \cite{Koenig:2011ti,Konig:2011nz} as well as two body systems with Neumann boundary conditions \cite{Tan:2012a}. 

\section{Formalism}

\begin{figure}
\centering
\includegraphics[width=0.4\textwidth]{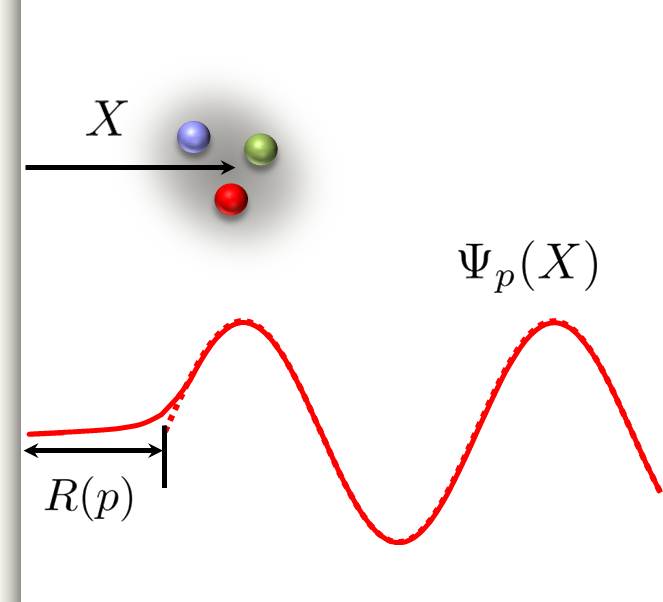}
\caption{Sketch showing the distance X from the wall to the center of mass of the bound state and the standing wave solution}
\label{COM}
\end{figure}

We consider a non-relativistic bound state in $d$ dimensions, with total mass
$M$, and arbitrary number of constituent particles.\ The bound state scatters
elastically off of a hard-wall boundary, implemented as a vanishing Dirichlet
boundary condition. We let $X$ be the distance from the wall to the center of
mass of the bound state. We work in the inertial frame where the only non-zero
component of the center-of-mass momentum is the component perpendicular to the
hard-wall boundary. Then we construct a standing wave solution with center-of-mass
momenta $\pm p$ perpendicular to the wall with reflection phase shift,
$\delta(p)$, and reflection radius, $R(p)=-\delta(p)/p$. The reflection radius
is a measure of the distance between the wall and the closest node of the
asymptotic standing wave, $\Psi_{p}(X)\propto\sin\left[  pX+\delta(p)\right]
$. This is shown in Fig. \ref{COM}.

Reflection in one dimension is analogous to $S$-wave scattering in three
dimensions and so we use the effective range expansion,%
\begin{equation}
p\cot\delta(p)=-\frac{1}{a_{R}}+\frac{1}{2}r_{R}p^{2}-\mathcal{P}_{R}%
p^{4}+\cdots, \label{ERE}%
\end{equation}
where $a_{R}$ is the scattering length, $r_{R}$ is the effective range, and
$\mathcal{P}_{R}$ is the shape parameter. \ Notice that at threshold, $a_{R}$
is equal to the reflection radius, $a_{R}=\lim_{p\rightarrow0}R(p)$. \ For a
completely rigid bound state, $R(p)=a_{R}$ for all $p$.\ \ However, we expect
that the bound state will not be perfectly rigid and so it will compress more
when the collision energy increases. This means that the reflection radius,
$R(p)$, will decrease with increasing bound state center-of-mass momentum,
$p$. The rate of decrease measures the compressibility of the bound state
under unilateral stress. 

We now consider the bound state confined to a $d$-dimensional cubic box with length $L$ and hard wall boundaries on all sides. \ In the limit of large $L$, we can approximate the wavefunction as a product of standing waves along each coordinate axis.  Each standing wave has half-wavelength equal to $L-2R\left[  p(L)\right]  $, and so the reflection radius can be related to the ground state confinement energy of the bound state as follows:

\begin{equation}
E(L)=\frac{p^{2}(L)\cdot d}{2M} \left[1 + O\left(  L^{-2}\right)\right], 
\label{energy_length}
\end{equation}
where
\begin{equation}
p(L)=\frac{\pi
}{L-2R\left[  p(L)\right]  }.%
\label{momentum_length}
\end{equation}

This relation can be used to determine the reflection radius as a function of the center-of-mass momentum.  The $O\left(L^{-2}\right)$ relative error in Eq.(\ref{energy_length}) arises because the different coordinates for $d > 1$ cannot be exactly separated as a product of standing waves.  This is due to the effects of double-wall collision near the wall intersections.  By wall intersections we mean the corners of a two-dimensional square or the edges of a three-dimensional cube.  In each case, the codimension of the wall intersections is two, and this explains the $O\left(L^{-2}\right)$ relative error.  In one dimension, however, there are no wall intersections, and so the error is exponentially small in $L$.

\subsection{General Two-body Bound States}

We now consider the more specific case of a bound state with zero orbital
angular momentum consisting of two distinguishable particles with masses
$m_{1}$ and $m_{2}$. The reduced mass, $\mu$, is defined in the usual way,
$\mu=(m_{1}^{-1}+m_{2}^{-1})^{-1}$. We let $E_{B}$ be the infinite volume binding
energy of the dimer, and $\kappa_{B}$ be the binding momentum defined by the
relation $E_{B}=-\kappa_{B}^{2}/(2\mu)$. \ $a_{B}$ is the $S$-wave scattering
length for shallow-binding particle-particle scattering. In the
shallow-binding limit, where $a_{B}$ is much larger than the range of the
interaction $\kappa_{B}=a_{B}^{-1}$, we can neglect the short-distance
physics. So, the reflection phase shift is a universal function of the
dimensionless ratio $p/\kappa_{B}$. \ In this paper we present the form of
this universal function.

\subsection{Shallow Dimer Reflection}

Next we use the principles of effective field theory to calculate the
reflection scattering length for a shallow dimer. While effective field theory is a well established method, in the current case we are dealing with a
non-homogenous system for which there is no immediately obvious small
expansion parameter. In the soft scattering limit, we use an adiabatic approximation for the center-of-mass motion of the system. We then perform an asymptotic expansion to account for the long-distance physics. The result is an expansion for $\kappa_{B}a_{R}$ with convergence controlled by an expansion parameter $e^{-2\kappa_{B}a_{R}}$. \ 

We let $\vec{r}_{1}$ and $\vec{r}_{2}$ be the coordinates of the two
constituent particles and assume an attractive short-range interaction given
by the Hamiltonian,%
\begin{equation}
H=-\frac{1}{2m_{1}}\vec{\nabla}_{r_{1}}^{2}-\frac{1}{2m_{2}}\vec{\nabla
}_{r_{2}}^{2}+C_{B}\bar{\delta}^{(d)}(\vec{r}_{1}-\vec{r}_{2}),
\end{equation}
where $\bar{\delta}^{(d)}$ is a regulated $d$-dimensional delta function.\ The
coefficient $C_{B}$ is tuned to produce a bound state with energy $E_{B}$ at
infinite volume. We take $\vec{r}=\vec{r}_{1}-\vec{r}_{2}$ to be the relative
separation between the particles. For any fixed center-of-mass coordinate, the
part of the Hamiltonian that is only dependent on the relative coordinate,
$\vec{r}$, is given by%
\begin{equation}
H_{\text{rel}}=-\frac{1}{2\mu}\vec{\nabla}_{r}^{2}+C_{B}\bar{\delta}%
^{(d)}(\vec{r}).
\end{equation}

\subsubsection{The Effective Hamiltonian}

Let $E_{K}$ be the kinetic energy of the moving dimer. To calculate the reflection scattering length it suffices to consider dimer-wall scattering in the limit $E_{K}\ll\left\vert E_{B}\right\vert $. In this low-energy limit we perform an adiabatic expansion for the center-of-mass motion. This technique is conceptually similar to the adiabatic hyperspherical approximation \cite{Macek:1968,Lin:1995a,Esry:1996a}.

We let $X$ be the distance from the wall to the center of mass of the dimer. We label the coordinate axes $1$, $2$,...,$d$, and take the $d^{th}$ dimension to be perpendicular to the wall. For fixed $X$
the hard-wall boundary at $(r_{1})_{d}=0$ gives a minimum value for the relative coordinate $r_{d}$,%

\begin{equation}
r_{d}^{\text{min}}=-\frac{M}{m_{2}}X.
\end{equation}
We now define $x_{-}$ as twice this distance,%
\begin{equation}
x_{-}(X)=-\frac{2M}{m_{2}}X.
\end{equation}
So $x_{-}\left(  X\right)  $ is twice the distance from the wall to $\vec
{r}_{2}$ when $\vec{r}_{1}$ touches the wall. \ In other words $\psi
_{X}(\vec{r})$ must vanish when $\vec{r}_{d}$ equals $-x_{-}(X)/2$. The factor of two simplifies the expansion to
be derived later. Similarly the hard-wall boundary for $\vec{r}_{2}$ gives a
maximum value for $r_{d}$,%
\begin{equation}
r_{d}^{\text{max}}=\frac{M}{m_{1}}X.
\end{equation}
As in the previous case, we define%
\begin{equation}
x_{+}(X)=\frac{2M}{m_{1}}X.
\end{equation}
So $x_{+}\left(  X\right)  $ is twice the distance from the wall to $\vec
{r}_{1}$ when $\vec{r}_{2}$ touches the wall. In other words $\psi
_{X}(\vec{r})$ must vanish when $\vec{r}_{d}$ equals $x_{+}(X)/2.$ Once again the factor of two has been added to
simplify the expansion to be derived later.

We now define the $d$-dimensional vectors%
\begin{equation}
\vec{r}_{-}(X)=(0,\cdots,0,x_{-}(X)),
\end{equation}%
\begin{equation}
\vec{r}_{+}(X)=(0,\cdots,0,x_{+}(X)).
\end{equation}
The magnitudes of the vectors are%
\begin{equation}
r_{-}(X)=\left\vert \vec{r}_{-}(X)\right\vert =-x_{-}(X),
\end{equation}%
\begin{equation}
r_{+}(X)=\left\vert \vec{r}_{+}(X)\right\vert =x_{+}(X).
\end{equation}

Let $E_{\kappa}$ be the kinetic energy of the moving dimer. For each $X$ it is only necessary to keep the ground state of $H_{\text{rel}}$ satisfying the hard-wall boundary condition. This is due to the fact that the contribution of the excited states are suppressed by powers of $E_{K}/\left\vert E_{B}\right\vert $ and hence cannot contribute to the reflection scattering length, but only to higher-order coefficients in the effective range expansion.

For each $X$ we take $\psi_{X}(\vec{r})$ to be the normalized ground state
wavefunction of $H_{\text{rel}}$ satisfying the hard-wall boundary constraint.
Then the eigenstates of the full Hamiltonian are%

\begin{equation}
\left\vert \Psi(X)\right\rangle \otimes\left\vert \psi_{X}(\vec{r})\right\rangle +O\left(  P^{2}/\kappa_{B}^{2}\right),
\end{equation}
where $P$ is the center-of-mass momentum of the dimer in the direction perpendicular to the wall. The low-energy effective Hamiltonian is%
\begin{equation}
H_{\text{eff}}=-\frac{1}{2M}\frac{\partial^{2}}{\partial X^{2}}+V(X)+T(X),
\end{equation}
where $V(X)$ is the adiabatic potential,%
\begin{equation}
V(X)=\left\langle \psi_{X}\right\vert H_{\text{rel}}\left\vert \psi
_{X}\right\rangle ,
\end{equation}
and $T(X)$ is the diagonal adiabatic correction, $T(X)=-\frac{1}%
{2M}\left\langle \psi_{X}\right\vert \frac{\partial^{2}}{\partial X^{2}%
}\left\vert \psi_{X}\right\rangle .$

Notice that the cross-term resulting from one derivative with respect to $X$
vanishes because of the fixed normalization of $\psi_{X}$. We use lattice
regularization to deal with the continuum limit singularity in the delta
function for $d>1.$

We define the local $X$-dependent energy, $E_{\vec{r}_{-},\vec{r}_{+}}$, and binding
momentum, $\kappa_{\vec{r}_{-},\vec{r}_{+},},$ in terms of the
adiabatic potential%
\begin{equation}
V(X)=\left\langle \psi_{X}\right\vert H_{\text{rel}}\left\vert \psi
_{X}\right\rangle = E_{\vec{r}_{+},\vec{r}_{-}} =-\frac{\kappa_{\vec{r}_{-},\vec{r}_{+}}^{2}}{2\mu},
\end{equation}
for the vanishing Dirichlet boundary condition. For large $X$ we can generate an asymptotic expansion for the local
$X$-dependent binding momentum, $\kappa_{\vec{r}_{-},\vec{r}_{+}},$ $V(X),$
and $T(X)$ in powers of $e^{-\kappa_{B}r_{+}\left(  X\right)  }$ and
$e^{-\kappa_{B}r_{-}\left(  X\right)  }$. The boundary conditions are
enforced by applying the method of images. 

\begin{figure}[htb]
\centering
\includegraphics[width=0.8\textwidth]{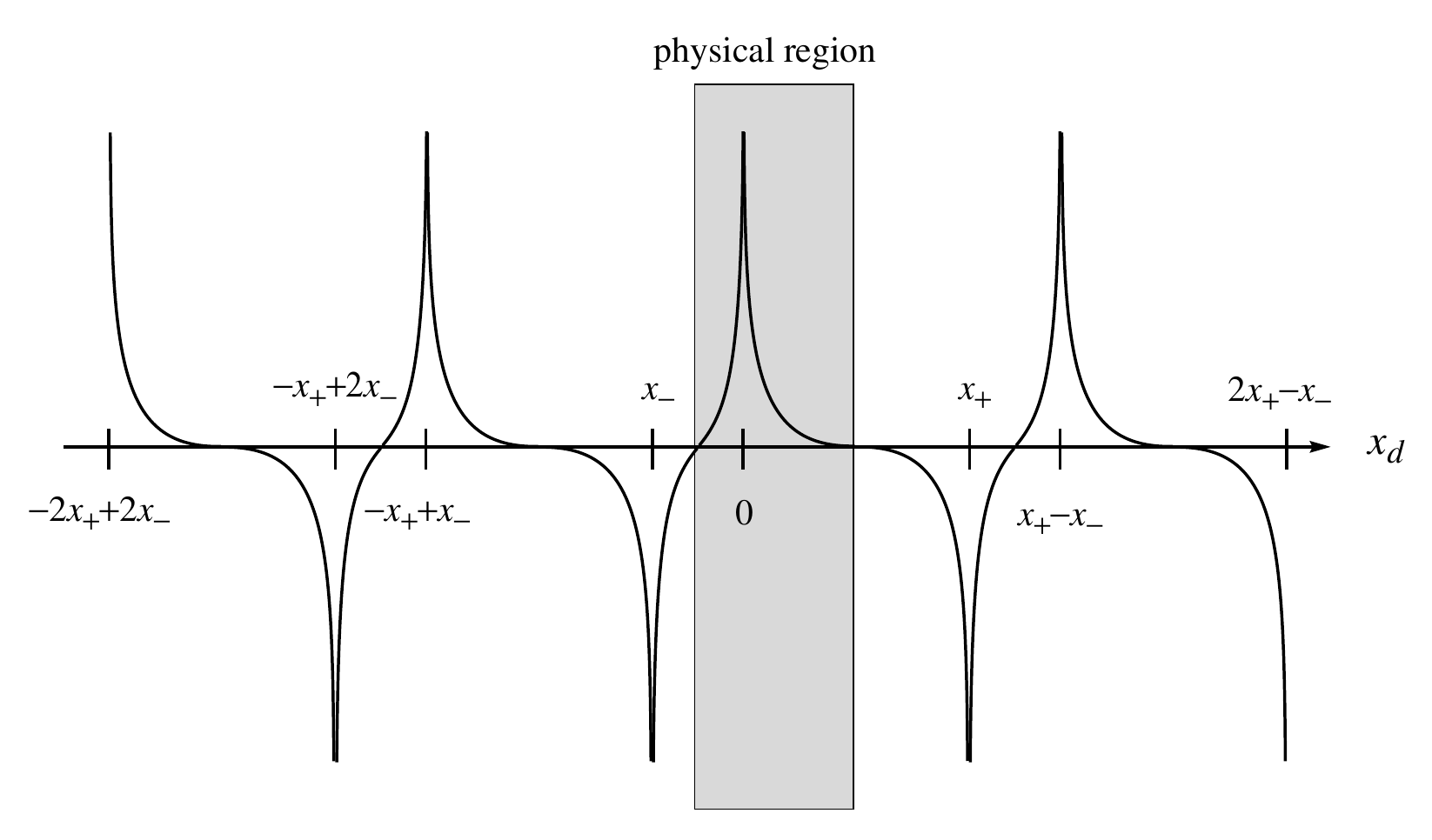}
\caption{Sketch of the wavefunction derived from the method of images}
\label{Psi}
\end{figure}

We consider two infinite periodic
chains of delta functions. The first chain includes delta
functions centered at multiples of $\vec{r}_{+}-$ $\vec{r}_{-}$. The second
chain includes a delta function centered at $\vec{r}_{+}$ plus multiples of
$\vec{r}_{+}-$ $\vec{r}_{-}$. We now construct a wavefunction which is the lowest energy eigenstate for this potential with alternating signs at the centers of the neighboring delta functions. For a picture of the wavefunction see Fig. \ref{Psi}.

For this wavefunction we use an ansatz that is a superposition of $d-$dimensional Yukawa functions.%

\begin{equation}
\psi_{\vec{r}_{-},\vec{r}_{+}}(\vec{r})=A_{\vec{r}_{-},\vec{r}_{+}}\phi
_{\vec{r}_{-},\vec{r}_{+}}(\vec{r}),\label{A}
\end{equation}
where%
\begin{align}
\phi_{\vec{r}_{-},\vec{r}_{+}}(\vec{r}) &  =y_{d,0}(\kappa_{\vec{r}_{-}%
,\vec{r}_{+}},\left\vert \vec{r}\right\vert )-y_{d,0}(\kappa_{\vec{r}_{-}%
,\vec{r}_{+}},\left\vert \vec{r}-\vec{r}_{-}\right\vert )-y_{d,0}(\kappa
_{\vec{r}_{-},\vec{r}_{+}},\left\vert \vec{r}-\vec{r}_{+}\right\vert
)\nonumber\\
&  +y_{d,0}(\kappa_{\vec{r}_{-},\vec{r}_{+}},\left\vert \vec{r}-\vec{r}%
_{+}+\vec{r}_{-}\right\vert )+y_{d,0}(\kappa_{\vec{r}_{-},\vec{r}_{+}},\vec
{r}+\vec{r}_{+}-\vec{r}_{-})+\cdots,\label{images}%
\end{align}
Here%
\begin{equation}
y_{d,0}(\kappa,r)=\int\frac{d^{d}\vec{p}}{\left(  2\pi\right)  ^{d}}%
\frac{e^{-i\vec{p}\cdot\vec{r}}}{p^{2}+\kappa^{2}}. \label{Yukawa}%
\end{equation}
is the generalized $d$ dimensional Yukawa function, and we also define%

\begin{equation}
f_{d}\left(  \kappa^{I},\kappa^{II},r\right)  =\int d^{d}\vec{r}^{\;\prime
}\;y_{d,0}(\kappa^{I},\left\vert \vec{r}^{\;\prime}\right\vert )y_{d,0}%
(\kappa^{II},\left\vert \vec{r}^{\;\prime}-\vec{r}\right\vert ),
\end{equation}
as the full overlap integral of two Yukawa functions in $d$ dimensions whose
centers are separated by a spatial distance $r$. In Eq.(\ref{A}) $A_{\vec{r}_{-},\vec{r}_{+}}$ is a function that normalizes $\psi_{\vec{r}%
_{-},\vec{r}_{+}}(\vec{r})$, and we find that%
\begin{align}
&  \int\frac{d^{d}\vec{p}}{\left(  2\pi\right)  ^{d}}\left[  \frac{1}%
{\frac{p^{2}}{2\mu}-E_{\vec{r}_{-},\vec{r}_{+}}}-\frac{1}{\frac{p^{2}}{2\mu
}-E_{B}}\right]  \nonumber\\
&  =\sum_{j}\int\frac{d^{d}\vec{k}}{\left(  2\pi\right)  ^{d}}\frac
{e^{-i\vec{p}\cdot(-(j+1)r_{+}+jr_{-})}}{\frac{p^{2}}{2\mu}-E_{\vec{r}%
_{-},\vec{r}_{+}}}-\sum_{j\neq0}\int\frac{d^{d}\vec{k}}{\left(  2\pi\right)
^{d}}\frac{e^{-i\vec{p}\cdot(-jr_{+}+jr_{-})}}{\frac{p^{2}}{2\mu}-E_{\vec
{r}_{-},\vec{r}_{+}}},
\end{align}
where $j$ is summed over integer values.

At first order in powers of $e^{-\kappa_{B}r_{+}\left(  X\right)  }$ and
$e^{-\kappa_{B}r_{-}\left(  X\right)  }$ we get a correction to the binding
momentum%
\begin{equation}
\kappa_{\vec{r}_{-},\vec{r}_{+}(1)}=\frac{\int\frac{d^{d}\vec{p}}{\left(
2\pi\right)  ^{d}}\left[  \frac{e^{i\vec{p}\cdot\vec{r}_{+}}}{p^{2}+\kappa
_{B}^{2}}+\frac{e^{i\vec{p}\cdot\vec{r}_{-}}}{p^{2}+\kappa_{B}^{2}}\right]
}{\frac{\partial}{\partial\kappa_{B}}\left[  \int\frac{d^{d}\vec{p}}{\left(
2\pi\right)  ^{d}}\frac{1}{p^{2}+\kappa_{B}^{2}}\right]  }.
\end{equation}
The first-order correction to the adiabatic potential is%
\begin{equation}
V^{\left(  1\right)  }(X)=-\frac{\kappa_{B}}{\mu}\kappa_{\vec{r}_{-},\vec
{r}_{+}(1)},
\end{equation}
and the first order diagonal adiabatic correction is%
\begin{align}
T^{\left(  1\right)  }(X)  &  =\frac{M}{2m_{1}^{2}f_{d}\left(  \kappa
_{B},\kappa_{B},0\right)  }y_{d,0}\left(  \kappa_{B},r_{+}\right) \nonumber\\
&  +\frac{M}{2m_{2}^{2}f_{d}\left(  \kappa_{B},\kappa_{B},0\right)  }%
y_{d,0}\left(  \kappa_{B},r_{-}\right)  .
\end{align}
So the first order correction to the
effective Hamiltonian is%

\begin{align}
V^{1}(X)+T^{1}(X)  &  = -\frac{\kappa_{B}}{\mu}\kappa_{\vec{r}_{-},\vec{r}_{+}%
(1)}+\frac{M}{2m_{1}^{2}f_{d}\left(  \kappa_{B},\kappa_{B},0\right)  }%
y_{d,0}\left(  \kappa_{B},r_{+}\right) \nonumber\\
&  +\frac{M}{2m_{2}^{2}f_{d}\left(  \kappa_{B},\kappa_{B},0\right)  }%
y_{d,0}\left(  \kappa_{B},r_{-}\right)  .
\end{align}

At second order the correction to the local binding momentum is%
\begin{align}
&  \kappa_{\vec{r}_{-},\vec{r}_{+}(2)}\nonumber\\
&  =\frac{1}{\frac{\partial}{\partial\kappa_{B}}y_{d,0}\left(  \kappa
_{B},0\right)  }\left\{  \frac{y_{d,0}\left(  \kappa_{B},r_{+}\right)
+y_{d,0}\left(  \kappa_{B},r_{-}\right)  }{\frac{\partial}{\partial\kappa_{B}%
}y_{d,0}\left(  \kappa_{B},0\right)  }\frac{\partial}{\partial\kappa_{B}%
}\left[  y_{d,0}\left(  \kappa_{B},r_{+}\right)  +y_{d,0}\left(  \kappa
_{B},r_{-}\right)  \right]  \right.  \nonumber\\
&  \left.  -2y_{d,0}\left(  \kappa_{B},r_{+}+r_{-}\right)  -\frac{\left[
y_{d,0}\left(  \kappa_{B},r_{+}\right)  +y_{d,0}\left(  \kappa_{B}%
,r_{-}\right)  \right]  ^{2}}{2\left[  \frac{\partial}{\partial\kappa_{B}%
}y_{d,0}\left(  \kappa_{B},0\right)  \right]  ^{2}}\frac{\partial^{2}%
}{\partial\kappa_{B}^{2}}y_{d,0}\left(  \kappa_{B},0\right)  \right\}  .
\end{align}
The correction to the adiabatic potential is%
\begin{equation}
V^{\left(  2\right)  }(X)=-\frac{\kappa_{B}}{\mu}\kappa_{\vec{r}_{-},\vec
{r}_{+}(2)}-\frac{1}{2\mu}\kappa_{\vec{r}_{-},\vec{r}_{+}(1)}^{2}.
\end{equation}
The diagonal adiabatic correction contains a number of terms which we unite as%
\begin{align}
T^{\left(  2\right)  }(X) &  =T_{(A)}^{(2)}(X)+T_{(B)}^{(2)}(X)+T_{(C)}%
^{(2)}(X)\nonumber\\
& +T_{(D1)}^{(2)}+T_{(D2)}^{(2)}(X)+T_{(D3)}^{(2)}(X)+T_{(D4)}^{(2)}(X)+T_{(E)}%
^{(2)}(X).\label{T_2}%
\end{align}

The first term in Eq.(\ref{T_2}) is%
\begin{equation}
T_{(A)}^{(2)}(X)=-\frac{f_{d}\left(  \kappa_{B},\kappa_{B},0\right)  }%
{2M}\left(  \frac{\partial A_{\vec{r}_{-},\vec{r}_{+}(1)}}{\partial
X}\right)  ^{2}+\cdots,\nonumber
\end{equation}
where the ellipses indicate terms at third order or higher. Here%

\begin{align}
A_{\vec{r}_{-},\vec{r}_{+}}  &  =\left[  f_{d}\left(  \kappa_{\vec{r}_{-}%
,\vec{r}_{+}},\kappa_{\vec{r}_{-},\vec{r}_{+}},0\right)  -f_{d}\left(
\kappa_{\vec{r}_{-},\vec{r}_{+}},\kappa_{\vec{r}_{-},\vec{r}_{+}}%
,r_{-}\right)  -f_{d}\left(  \kappa_{\vec{r}_{-},\vec{r}_{+}},\kappa_{\vec
{r}_{-},\vec{r}_{+}},r_{+}\right)  \right. \nonumber\\
&  \left.  +2f_{d}\left(  \kappa_{\vec{r}_{-},\vec{r}_{+}},\kappa_{\vec{r}%
_{-},\vec{r}_{+}},r_{+}+r_{-}\right)  +\cdots\right]  ^{-1/2},
\end{align}
arises from the normalization of the relative coordinate wavefunction. Once
again the ellipses indicate third order or higher terms. For more details see
Appendix \ref{Diagonal}. The derivative of the normalization function, at
first order, with respect to $X$ is
\begin{align}
&  \frac{\partial A_{\vec{r}_{-},\vec{r}_{+}(1)}}{\partial X}\nonumber\\
&  =\frac{1}{2}\frac{\frac{2M}{m_{1}}\frac{\partial}{\partial r_{+}}%
f_{d}\left(  \kappa_{B},\kappa_{B},r_{+}\right)  +\frac{2M}{m_{2}}%
\frac{\partial}{\partial r_{-}}f_{d}\left(  \kappa_{B},\kappa_{B}%
,r_{-}\right)  -\frac{\partial\kappa_{\vec{r}_{-},\vec{r}_{+}(1)}}{\partial
X}\frac{\partial}{\partial\kappa_{B}}f_{d}\left(  \kappa_{B},\kappa
_{B},0\right)  }{\left[  f_{d}\left(  \kappa_{B},\kappa_{B},0\right)  \right]
^{3/2}}. \label{dA_dx_1}%
\end{align}
The derivative of the first order correction to the binding energy with
respect to $X$ is%
\begin{equation}
\frac{\partial\kappa_{\vec{r}_{-},\vec{r}_{+}(1)}}{\partial X}=\frac
{\frac{2M}{m_{1}}\frac{\partial}{\partial r_{+}}y_{d,0}\left(  \kappa
_{B},r_{+}\right)  +\frac{2M}{m_{2}}\frac{\partial}{\partial r_{-}}%
y_{d,0}\left(  \kappa_{B},r_{-}\right)  }{\frac{\partial}{\partial\kappa_{B}%
}y_{d,0}\left(  \kappa_{B},0\right)  }.
\end{equation}

The second term in Eq.(\ref{T_2}), due to two derivatives with respect to the
spatial coordinates without any contribution from the first-order corrections
$A_{\vec{r}_{-},\vec{r}_{+}(1)}$ and $\kappa_{\vec{r}_{-},\vec{r}_{+}(1)}$, is

\begin{align}
T_{(B)}^{(2)}(X)  &  =-\frac{M}{\mu^{2}f_{d}\left(  \kappa_{B},\kappa
_{B},0\right)  }y_{d,0}(\kappa_{B},r_{+}+r_{-})\nonumber\\
&  +\frac{r_{+}+r_{-}}{\mu f_{d}\left(  \kappa_{B},\kappa_{B},0\right)
}y_{d,-1}(\kappa_{B},r_{+}+r_{-}),
\end{align}

The third term in Eq.(\ref{T_2}), proportional to $\kappa_{\vec{r}_{-}%
,\vec{r}_{+}(1)}$ and $A_{\vec{r}_{-},\vec{r}_{+}(1)}$, is%
\begin{align}
T_{(C)}^{(2)}(X)  &  =\frac{M\kappa_{\vec{r}_{-},\vec{r}_{+}(1)}}{2m_{1}%
^{2}f_{d}\left(  \kappa,\kappa,0\right)  }\left.  \frac{\partial}%
{\partial\kappa}\left[  y_{d,0}\left(  \kappa,r_{+}\right)  \right]
\right\vert _{\kappa=\kappa_{B}}+\frac{M\kappa_{\vec{r}_{-},\vec{r}_{+}(1)}%
}{2m_{2}^{2}f_{d}\left(  \kappa,\kappa,0\right)  }\left.  \frac{\partial
}{\partial\kappa}\left[  y_{d,0}\left(  \kappa,r_{-}\right)  \right]
\right\vert _{\kappa=\kappa_{B}}\nonumber\\
& +\frac{A_{\vec{r}_{-},\vec{r}_{+}(1)}}{\left[  f_{d}\left(  \kappa
_{B},\kappa_{B},0\right)  \right]  ^{1/2}}\left[  \frac{M}{m_{1}^{2}}%
y_{d,0}\left(  \kappa_{B},r_{+}\right)  +\frac{M}{m_{2}^{2}}y_{d,0}\left(
\kappa_{B},r_{-}\right)  \right] \label{TC_2},
\end{align}%
where $y_{d,1}\left(  \kappa,r\right)  ,$ $y_{d,2}\left(  \kappa,r\right)  ,$
and $y_{d,3}\left(  \kappa,r\right)  $ are the generalized Yukawa functions
defined later, in Eqs.~(\ref{Yukawa_1D}), (\ref{Yukawa_2D}), and (\ref{Yukawa_3D}), for $d=1$, $2$, and $3$. For the terms in the second line of Eq.(\ref{T_2}), we find%

\begin{align}
T_{(D1)}^{(2)}(X)  &  =-\frac{\kappa_{B}}{m_{1}f_{d}\left(  \kappa_{B}%
,\kappa_{B},0\right)  }\frac{\partial\kappa_{\vec{r}_{-},\vec{r}_{+}(1)}%
}{\partial X}\left\{  \frac{r_{+}^{2}}{8}y_{d,1}\left(  \kappa_{B}%
,r_{+}\right)  \right. \nonumber\\
&  \left.  +\frac{r_{+}}{2}y_{d,2}\left(  \kappa_{B},r_{+}\right)  +\frac
{3}{4}y_{d,3}\left(  \kappa_{B},r_{+}\right)  \right\}  ,
\end{align}

\begin{align}
T_{(D2)}^{(2)}(X)  &  =-\frac{\kappa_{B}}{m_{2}f_{d}\left(  \kappa_{B}%
,\kappa_{B},0\right)  }\frac{\partial\kappa_{\vec{r}_{-},\vec{r}_{+}(1)}%
}{\partial X}\left\{  \frac{r_{-}^{2}}{8}y_{d,1}\left(  \kappa_{B}%
,r_{-}\right)  \right. \nonumber\\
&  \left.  +\frac{r_{-}}{2}y_{d,2}\left(  \kappa_{B},r_{-}\right)  +\frac
{3}{4}y_{d,3}\left(  \kappa_{B},r_{-}\right)  \right\}  ,
\end{align}

\begin{equation}
T_{(D3)}^{(2)}(X)=\frac{\kappa_{B}}{4m_{1}f_{d}\left(  \kappa_{B},\kappa
_{B},0\right)  }\frac{\partial\kappa_{\vec{r}_{-},\vec{r}_{+}(1)}}{\partial
X}\left[  r_{+}y_{d,2}\left(  \kappa_{B},r_{+}\right)  +3y_{d,3}\left(
\kappa_{B},r_{+}\right)  \right]  , \label{TD3_2}%
\end{equation}

\begin{equation}
T_{(D4)}^{(2)}(X)=\frac{\kappa_{B}}{4m_{2}f_{d}\left(  \kappa_{B},\kappa
_{B},0\right)  }\frac{\partial\kappa_{\vec{r}_{-},\vec{r}_{+}(1)}}{\partial
X}\left[  r_{-}y_{d,2}\left(  \kappa,r_{-}\right)  +3y_{d,3}\left(
\kappa,r_{-}\right)  \right]  , \label{TD4_2}%
\end{equation}

\begin{align}
T_{(E)}^{(2)}(X)  &  =\frac{\left(  \frac{\partial\kappa_{\vec{r}_{-},\vec
{r}_{+}(1)}}{\partial X}\right)  ^{2}}{24M\kappa_{B}^{3}f_{d}\left(
\kappa_{B},\kappa_{B},0\right)  }\left\{  -3\frac{\partial y_{d,0}}%
{\partial\kappa_{B}}(\kappa_{B},0)\right. \nonumber\\
&  \left.  +3\kappa\frac{\partial^{2}y_{d,0}}{\partial\kappa_{B}^{2}}%
(\kappa_{B},0)-\kappa^{2}\frac{\partial^{3}y_{d,0}}{\partial\kappa_{B}^{3}%
}(\kappa_{B},0)\right\}  ,
\end{align}
and the effective Hamiltonian is%

\begin{equation}
H_{\text{eff}}^{\left(  2\right)  }=-\frac{1}{2M}\frac{\partial^{2}}{\partial
X^{2}}-\frac{\kappa_{B}}{\mu}\kappa_{\vec{r}_{-},\vec{r}_{+}(2)}-\frac
{1}{2\mu}\kappa_{\vec{r}_{-},\vec{r}_{+}(1)}^{2}+T^{(2)}(X),
\end{equation}
where $\kappa_{\vec{r}_{-},\vec{r}_{+}(1)},$ $\kappa_{\vec{r}_{-},\vec{r}%
_{+}(2)},$\ and $T^{(2)}(X)$ are as defined above.

\section{ Results}

In this section we present the results of our effective field theory expansions
in one, two, and three dimensions. Using the definitions given below for the
generalized Yukawa function, full two Yukawa function overlap integral, and effective Hamiltonia terms from the previous section, we present the
expressions for the effective Hamiltonian in one, two, and three dimensions.
In all cases at zeroth order we start with the infinite volume result,

\begin{equation}
H_{\text{eff}}^{\left(0\right)} = -\frac{1}{2M}\frac{\partial^{2}}{\partial
X^{2}}+V_{X}^{\left(0\right)}+T_{X}^{\left(0\right)},
\end{equation}
where
\begin{equation}
V_{X}^{\left(0\right)}+T_{X}^{\left(0\right)}=E_{B}.
\end{equation}

\subsection{Results in One Dimension}

The generalized Yukawa function in one dimension is%
\begin{equation}
y_{1,0}\left(  \kappa,r\right)  =\frac{e^{-\kappa r}}{2\kappa},
\label{Yukawa_1D}%
\end{equation}
and the full two Yukawa function overlap integral in one dimension is%

\begin{equation}
f_{1}\left(  \kappa,\kappa,r\right)  =\frac{e^{-\kappa r}}{4\kappa^{3}}\left(
1+\kappa r\right)  . \label{Overlap_1D}%
\end{equation}

\subsubsection{First Order Effective Hamiltonian in One
Dimension}

At first order the effective Hamiltonian for the one-dimensional case is%

\begin{equation}
H_{\text{eff}}^{\left(  1\right)  }=H_{\text{eff}}^{\left(0\right)}+V_{X}^{\left(1\right)}+T_{X}^{\left(1\right)} ,
\end{equation}

where the effective potential is%

\begin{equation}
V^{(1)}(X)+T^{(1)}(X)=\frac{\kappa_{B}^{2}M^{2}}{m_{1}m_{2}}\left[
\frac{e^{-\kappa_{B}r_{+}\left(  X\right)  }}{m_{1}}+\frac{e^{-\kappa_{B}%
r_{-}\left(  X\right)  }}{m_{2}}\right]  .\label{V1_1D}%
\end{equation}

\subsubsection{Second Order Effective Hamiltonian in One Dimension}

At second order in one dimension the effective Hamiltonian is%

\begin{equation}
H_{\text{eff}}^{\left(  2\right)  }=H_{\text{eff}}^{\left( 1\right)}+T^{(2)}(X)+V^{(2)}(X),
\end{equation}
where%

\begin{equation}
V^{(2)}(X)=\frac{M\kappa_{B}^{2}}{2m_{1}m_{2}}\left(  V_{-}^{(2)}%
(X)+V_{+}^{(2)}(X)+V_{(A)}^{(2)}(X)\right),
\end{equation}

\begin{equation}
V_{\pm}^{(2)}(X)=e^{-2\kappa_{B}r_{\pm}}(2\kappa_{B}r_{\pm}-1),
\end{equation}

\begin{equation}
V_{(A)}^{(2)}(X)=e^{-\kappa_{B}(r_{+}+r_{-})}\left(  \kappa_{B}r_{+}%
+\kappa_{B}r_{-}-3\right)
\end{equation}
and%

\begin{align}
T^{\left(  2\right)  }(X) &  =T_{(A)}^{(2)}(X)+T_{(B)}^{(2)}(X)+T_{(C)}%
^{(2)}(X)\nonumber\\
&  +T_{(D1)}^{(2)}(X)+T_{(D2)}^{(2)}(X)+T_{(D3)}^{(2)}(X)+T_{(D4)}^{(2)}(X)+T_{(E)}^{(2)}(X).
\end{align}
For the definitions of $T_{(A)}^{(2)},$ $T_{(B)}^{(2)},$ $T_{(C)}^{(2)},$
$T_{(D1-D4)}^{(2)},$ and
$T_{(E)}^{(2)}$, see Appendix \ref{Diagonal1D}.%

%TCIMACRO{\FRAME{ftbpFU}{4.0465in}{1.887in}{0pt}{\Qcb{Plot of the effective
%potential at zeroth, first, and second order in one spatial dimension for mass
%ratio $m_{2}/m_{1}=4$.}}{\Qlb{1DPlot}}{hardwalleffpotential1d.eps}%
%width 4.0465in;  height 1.887in;  depth 0pt;  original-width 3.9972in;
%original-height 1.8498in;  cropleft "0";  croptop "1";  cropright "1";
%filename 'HardWallEffPotential1D.eps';file-properties "XNPEU";}} }%
%BeginExpansion
\begin{figure}[ht]
\centering
\includegraphics[width=0.8\textwidth]{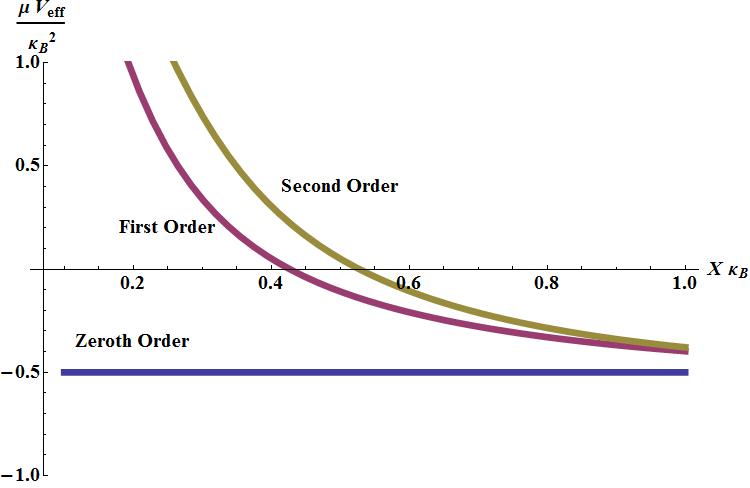}
\caption{Plot of the effective potential at zeroth, first, and second order in
one spatial dimension for mass ratio $m_{2}/m_{1}=4$.}
\label{1DPlot}
\end{figure}
%EndExpansion

Fig. \ref{1DPlot} shows a plot of the effective potential, at zeroth, first,
and second order, in one spatial dimension for mass ratio $m_{2}/m_{1}=4$. The
effective potential is plotted in dimensionless units of $X\kappa_{B}$ along
the horizontal axis and $\frac{\mu V_{eff}}{\kappa_{B}^{2}}$ along the
vertical axis. We see that the effective potential crosses zero at $X\kappa_{B}\sim0.5$, and the convergence for the expansion is good for $X\kappa_{B}\gtrsim0.5$. This part of the potential gives the dominant contribution to the reflection scattering length. 

\subsection{Results in Two Dimensions}

The generalized Yukawa function in two dimensions is%
\begin{equation}
y_{2,0}\left(  \kappa,r\right)  =\frac{1}{2\pi}K_{0}(\kappa r),
\label{Yukawa_2D}
\end{equation}
and the full two Yukawa function overlap integral in two dimensions is%
\begin{equation}
f_{2}\left(  \kappa,\kappa,r\right)  =\frac{r}{4\pi\kappa}K_{1}(\kappa r),
\end{equation}
where $K_{n}$ is the modified Bessel function of the $2^{nd}$ kind.

\subsubsection{First Order Effective Hamiltonian in Two Dimensions}

In two dimensions the first order effective Hamiltonian is%

\begin{align}
H_{\text{eff}}^{\left(  1\right)  }  &  =H_{\text{eff}}^{\left(  0\right)  }+\left(  \frac{1}{\mu}+\frac{M}{m_{1}^{2}}\right)  \kappa
_{B}^{2}K_{0}\left[  \kappa_{B}r_{+}(X)\right] \nonumber\label{d2_order1}\\
&  +\left(  \frac{1}{\mu}+\frac{M}{m_{2}^{2}}\right)  \kappa_{B}^{2}%
K_{0}\left[  \kappa_{B}r_{-}(X)\right]  ,
\end{align}
where the adiabatic potential is%

\begin{equation}
V^{(1)}(X)=\frac{\kappa_{B}^{2}}{\mu}\left[  K_{0}(\kappa_{B}r_{+}%
)+K_{0}(\kappa_{B}r_{-})\right]  ,
\end{equation}
 and the diagonal adiabatic correction is%

\begin{equation}
T^{\left(  1\right)  }(X)=\frac{\kappa_{B}^{2}M}{m_{1}^{2}}K_{0}(\kappa
_{B}r_{+})+\frac{\kappa_{B}^{2}M}{m_{2}^{2}}K_{0}(\kappa_{B}r_{-}).
\end{equation}

\subsubsection{Second Order Effective Hamiltonian in Two Dimensions}

At second order in two dimensions we find the adiabatic potential%

\begin{equation}
V^{(2)}(X)=-\frac{\kappa_{B}}{\mu}\kappa_{\vec{r}_{-},\vec{r}_{+}(2)}%
-\frac{1}{2\mu}\kappa_{\vec{r}_{-},\vec{r}_{+}(1)}^{2}.%
\end{equation}
The diagonal adiabatic correction is%

\begin{align}
T^{\left(  2\right)  }(X) &  =T_{(A)}^{(2)}(X)+T_{(B)}^{(2)}(X)+T_{(C)}%
^{(2)}(X)\nonumber\\
&  +T_{(D1)}^{(2)}(X)+T_{(D2)}^{(2)}(X)+T_{(D3)}^{(2)}(X)+T_{(D4)}^{(2)}(X)+T_{(E)}^{(2)}(X).
\end{align}
For the definitions of $\kappa_{\vec{r}_{-},\vec{r}_{+}(1)},$ $\kappa
_{\vec{r}_{-},\vec{r}_{+}(2)},$ $T_{(A)}^{(2)},$ $T_{(B)}^{(2)},$
$T_{(C)}^{(2)},$ $T_{(D1-D4)}^{(2)},$ and $T_{(E)}^{(2)}$, see Appendix \ref{Diagonal2D}.%

%TCIMACRO{\FRAME{ftbpFU}{4.0465in}{1.887in}{0pt}{\Qcb{Plot of the effective
%potential at zeroth, first, and second order in two spatial dimensions for
%mass ratio $m_{2}/m_{1}=4.$}}{\Qlb{2DPlot}}{hardwall_potential_2d.eps}%
%width 4.0465in;  height 1.887in;  depth 0pt;  original-width 3.9972in;
%original-height 1.8498in;  cropleft "0";  croptop "1";  cropright "1";
%filename 'HardWall_Potential_2D.eps';file-properties "XNPEU";}} }%
%BeginExpansion
\begin{figure}
\centering
\includegraphics[width=0.8\textwidth]{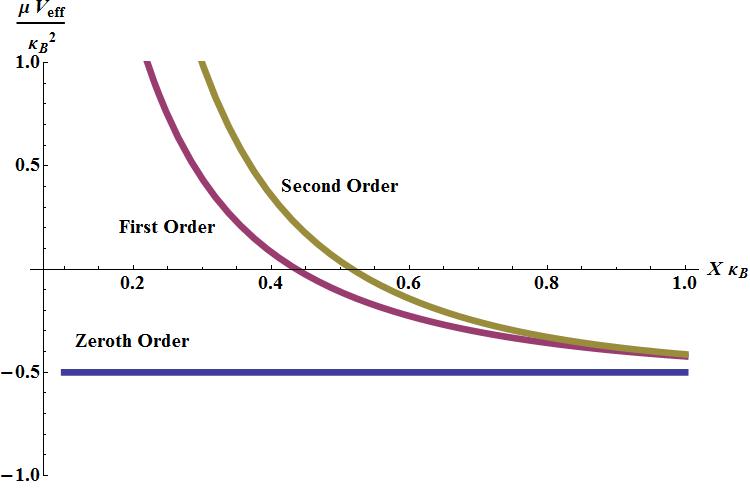}
\caption{Plot of the effective potential at zeroth, first, and second order in
two spatial dimensions for mass ratio $m_{2}/m_{1}=4.$}
\label{2DPlot}
\end{figure}
%EndExpansion

Fig. \ref{2DPlot} shows a plot of the effective potential in two spatial
dimensions for mass ratio $m_{2}/m_{1}=4.$ The effective potential
is plotted in dimensionless units of $X\kappa_{B}$ along the horizontal axis
and $\frac{\mu V_{eff}}{\kappa_{B}^{2}}$ along the vertical axis. Again the effective potential crosses zero at $X\kappa_{B}\sim0.5$, and the convergence of the expansion is good for larger values of $X\kappa_{B}$.

\subsection{Results in Three Dimensions}

The generalized Yukawa function in three dimensions is%
\begin{equation}
y_{3,0}\left(  \kappa,r\right)  =\frac{e^{-\kappa r}}{4\pi r},%
\label{Yukawa_3D}
\end{equation}
and the full two Yukawa function overlap integral in three dimensions is%
\begin{equation}
f_{3}\left(  \kappa,\kappa,r\right)  =\frac{e^{-\kappa r}}{8\pi\kappa}.
\end{equation}

\subsubsection{First Order Effective Hamiltonian in Three
Dimensions}

In three dimensions at first order the effective Hamiltonian is%

\begin{equation}
H_{\text{eff}}^{\left(  1\right)  }=H_{\text{eff}}^{\left(  0\right)  } +T^{\left(  1\right)  }(X)+ V^{\left(  1\right)  }(X)  ,
\end{equation}
where the effective potential is%

\begin{equation}
T^{(1)}(X)+V^{(1)}(X)=\frac{\kappa_{B}M}{2m_{1}m_{2}X}\left[  e^{-\kappa
_{B}r_{+}\left(  X\right)  }+e^{-\kappa_{B}r_{-}\left(  X\right)  }\right]
.\label{V1_3D}%
\end{equation}

\subsubsection{Second Order Effective Hamiltonian in Three
Dimensions}

At second order in three dimensions we find that the adiabatic potential is%

\begin{equation}
V^{\left(  2\right)  }(X)=-\frac{\kappa_{B}}{\mu}\kappa_{\vec{r}_{-},\vec
{r}_{+}(2)}-\frac{1}{2\mu}\kappa_{\vec{r}_{-},\vec{r}_{+}(1)}^{2}.%
\end{equation}
For the diagonal adiabatic correction, we find%

\begin{align}
T^{\left(  2\right)  }(X) &  =T_{(A)}^{(2)}(X)+T_{(B)}^{(2)}(X)+T_{(C)}%
^{(2)}(X)\nonumber\\
&  +T_{(D1)}^{(2)}(X)+T_{(D2)}^{(2)}(X)+T_{(D3)}^{(2)}(X)+T_{(D4)}^{(2)}(X)+T_{(E)}^{(2)}(X),
\end{align}

For the definitions of $\kappa_{\vec{r}_{-},\vec{r}_{+}(1)},$ $\kappa
_{\vec{r}_{-},\vec{r}_{+}(2)},$ $T_{(A)}^{(2)},$ $T_{(B)}^{(2)},$
$T_{(C)}^{(2)},$ $T_{(D1-D4)}^{(2)},$  and $T_{(E)}^{(2)}$, see Appendix \ref{Diagonal3D}.%

%TCIMACRO{\FRAME{ftbpFU}{4.0465in}{1.887in}{0pt}{\Qcb{Plot of the effective
%potential at zeroth, first, and second order in three spatial dimensions for
%mass ratio $m_{2}/m_{1}=4.$}}{\Qlb{3DPlot}}{hardwall_potential_3d.eps}%
%width 4.0465in;  height 1.887in;  depth 0pt;  original-width 3.9972in;
%original-height 1.8498in;  cropleft "0";  croptop "1";  cropright "1";
%filename 'HardWall_Potential_3D.eps';file-properties "XNPEU";}} }%
%BeginExpansion
\begin{figure}[ht]
\centering
\includegraphics[width=0.8\textwidth]{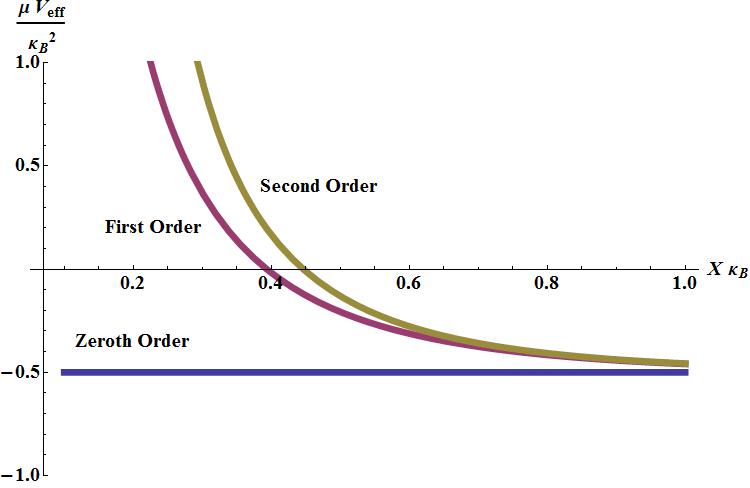}
\caption{Plot of the effective potential at zeroth, first, and second order in
three spatial dimensions for mass ratio $m_{2}/m_{1}=4.$}
\label{3DPlot}
\end{figure}
%EndExpansion

Fig. \ref{3DPlot} shows a plot of the effective potential in three spatial
dimensions for mass ratio $m_{2}/m_{1}=4.$ As in the one and two dimensional
cases, the effective potential is plotted in dimensionless units of
$X\kappa_{B}$ along the horizontal axis and $\frac{\mu V_{eff}}{\kappa_{B}%
^{2}}$ along the vertical axis. The effective potential crosses the horizontal axis at $X\kappa_{B}\sim0.4$, and the convergence of the expansion is good for $X\kappa_{B}\gtrsim0.4$.

\section{Scattering Length and Reflection Radius}

From these effective potentials it is straightforward to compute the reflection scattering length up to second order. This process can be
carried forward to any order. The net result is an expansion with an expansion
parameter of size $e^{-\kappa_{B}r_{\pm}(a_{R})}$ $\leq$ $e^{-2\kappa_{B}%
a_{R}}$.\ The larger $\kappa_{B}a_{R}$, the faster the convergence of the
expansion.\ First- and second-order results for the one-dimensional system are
shown in Fig.~\ref{latt_1D}. Results for the two-dimensional system are shown
in Fig. \ref{latt_2D}, and results for the three-dimensional system are shown
in Fig.~\ref{latt_3D}.\ In each case the agreement with lattice results is
consistent with third-order corrections of size $\leq$ $e^{-6\kappa_{B}a_{R}}$.

\subsection{Numerical Results}

We have calculated the dimer-wall reflection phase shift using Hamiltonian
lattice methods in one, two, and three dimensions. We implement the
two-particle interaction as an attractive point-like interaction. The lattice Hamiltonian in $d$ dimensions is
given by

\begin{align}
H &  =\frac{d}{m}\sum_{i=1,2}\sum_{\vec{n}}a_{i}^{\dagger}\left(  \vec
{n}\right)  a_{i}\left(  \vec{n}\right)  \nonumber\\
&  -\frac{1}{2m}\sum_{i=1,2}\sum_{\vec{n}}\sum_{l=1}^{d}\left[  a_{i}%
^{\dagger}\left(  \vec{n}\right)  a_{i}\left(  \vec{n}+\hat{l}\right)
+a_{i}^{\dagger}\left(  \vec{n}\right)  a_{i}\left(  \vec{n}-\hat{l}\right)
\right]  \nonumber\\
&  +c\sum_{\vec{n}}a_{1}^{\dagger}\left(  \vec{n}\right)
a_{1}\left(  \vec{n}\right)  a_{2}^{\dagger}\left(  \vec{n}\right)
a_{2}\left(  \vec{n}\right)  ,\label{Lattice_Hamiltonian}%
\end{align}
where $a_{i}^{\dagger}$ and $a_{i}$ are the creation and annihilation operators for particle $i$, 
$\vec{n}$ labels the lattice site, and $\hat{l}$ is the unit vector in the $l^{th}$ direction.
We take parameters and operators to be dimensionless by multiplying physical values by the
appropriate power of the lattice spacing.

Two parallel hard-wall boundaries are spaced a distance $L_{z}$ apart. We impose
periodic boundary conditions in the perpendicular directions with length $L$, and consider dimer states where the momentum is 
perpendicular to the wall. We calculate\ the confinement energy as a function of 
$L_{z}$ using sparse-matrix eigenvector methods and from this calculation determine 
the reflection phase shift using Eq.~(\ref{energy_length}).\ For various mass ratios of the two particles, we repeat our calculations using 
successively smaller lattice spacings. This allows us to extrapolate the reflection radius and dimer kinetic energy to the continuum limit and 
determine universal results in the shallow-binding limit.\ For the two- and three-dimensional 
cases we also perform infinite volume extrapolations in the dimension perpendicular 
to the wall. 

The coefficient of the delta function interaction in Eq.~(\ref{Lattice_Hamiltonian}) is $c$. In order to take the continuum limit in one dimension, we fix $L_{z}|c|$ and extrapolate both 
$R\kappa_{B}$ and $E_{K}/|E_{B}|$ for three different values of $L$. The values of $L_{z}|c|$ we consider are $L_{z}|c|=8,10,20$. For $L_{z}|c|=8$, we calculate the ground state values for $R\kappa_{B}$ and $E_{K}/|E_{B}|$ for $L_{z}=80,160,240$. For $L_{z}|c|=10$, we consider the ground state and first two excited states for $L_{z}=100,200,300$. For $L_{z}|c|=20$, we consider the ground state for $L_{z}=200,300,400$.   

In two and three dimensions, we first take the infinite volume limit in the perpendicular directions by extrapolating $L\to\infty$. We then extrapolate to the continuum limit using several values for $L_{z}$ while keeping $\kappa_{B}L_{z}$ fixed, where $\kappa_{B}$ is the dimer binding momentum.
In two dimensions we consider the ground state and first excited state for $L_{z}\kappa_{B}=6.8,10.2$. For $L_{z}\kappa_{B}=6.8$ we perform calculations using $L_{z}=40,60,80,100$, and for $L_{z}\kappa_{B}=10.2$ we consider $L_{z}=60,90,120,150$. 

In three dimensions we consider the ground state and first excited state for $L_{z}\kappa_{B}=4.9,6.0$. For $L_{z}\kappa_{B}=4.9$ we perform calculations using $L_{z}=20,30,40,$ and for $L_{z}\kappa_{B}=6.0$ we consider $L_{z}=24,36,48$. 
For a summary of the continuum extrapolations performed, including the lattice volumes used, see Appendix \ref{Extrap}.

\subsubsection{Numerical Results in One Dimension}

Fig.~\ref{latt_1D} shows the one-dimensional results in the shallow-binding
limit, plotted as the reflection radius versus dimer kinetic energy, $E_{K}$,
in dimensionless units, $\kappa_{B}R(E_{K})$ versus $E_{K}/\left\vert
E_{B}\right\vert $. The results are plotted at several different mass ratios,
$m_{2}/m_{1}$.\ As expected the reflection radius decreases monotonically with increasing
energy.\ One interesting feature is the dependence on the mass ratio.\ For
larger $m_{2}/m_{1}$ the reflection radius is larger but decreases quickly with increasing
energy. This indicates that a bound state with constituent particle masses that
are very different will behave like a large but soft deformable ball upon contact with the wall boundary.

When $m_{2}/m_{1}=1$ in one dimension the problem is integrable and exactly
solvable via the Bethe Ansatz.\ From the Bethe Ansatz we get%
\begin{equation}
p\cot\delta(p)=-2\kappa_{B}\text{,} \label{ERE_BA}%
\end{equation}%
\begin{equation}
\kappa_{B}R(E_{K})=\frac{1}{2}\sqrt{\frac{\left\vert E_{B}\right\vert }{E_{K}%
}}\tan^{-1}\sqrt{\frac{E_{K}}{\left\vert E_{B}\right\vert }}. \label{R_BA}%
\end{equation}
{\
%TCIMACRO{\FRAME{ftbpFU}{2.2675in}{2.2278in}{0pt}{\Qcb{One-dimensional lattice
%results for the reflection radius versus dimer kinetic energy in the
%shallow-binding limit. Also shown are Bethe Ansatz results for $m_{2}/m_{1}=1$
%and first- and second-order results for the expansion of $\kappa_{B}a_{R}$
%described later in the text.}}{\Qlb{latt_1D}}{radius_vs_e_1d.eps}%
%width 2.2675in;  height 2.2278in;  depth 0pt;  original-width 5.5988in;
%original-height 5.4993in;  cropleft "0";  croptop "1";  cropright "1";
%filename 'HardwallPaper/hardwallPaperSept/radius_vs_E_1D.eps';file-properties "XNPEU";}%
%} }%
%BeginExpansion
\begin{figure}[ht]
\centering
\includegraphics[width=0.8\textwidth]{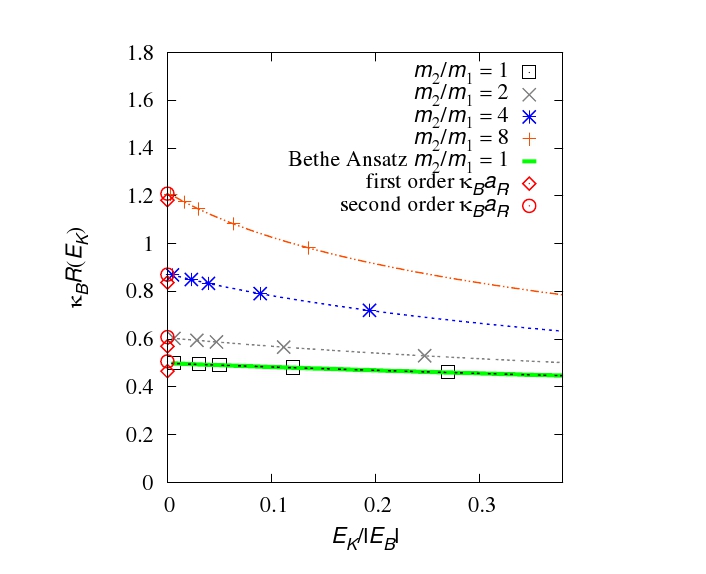}
\caption{One-dimensional lattice results for the reflection radius versus
dimer kinetic energy in the shallow-binding limit. Also shown are Bethe Ansatz
results for $m_{2}/m_{1}=1$ and first- and second-order results for the
expansion of $\kappa_{B}a_{R}$ described in the text.}
\label{latt_1D}
\end{figure}
%EndExpansion
As shown in }Fig.~\ref{latt_1D}, the lattice results agree with the solution
given by the Bethe Ansatz.\ Fig.~\ref{latt_1D} also shows the first- and
second-order analytic results for the expansion of $\kappa_{B}a_{R}$ derived
from our general effective theory.

Table~\ref{1D_table} presents the coefficients of the effective range
expansion for one-dimensional dimer-wall reflection.\ The error estimates are
from the least-squares fitting used in the lattice extrapolation and the effective
range expansion. From Eq.~(\ref{ERE_BA}), we see that the Bethe Ansatz gives
$\kappa_{B}a_{R}=1/2$ for $m_{2}/m_{1}=1$, with all other coefficients equal
to zero. This completely agrees with the lattice results.\ These results are
universal and could be verified using other theoretical methods or experiments
such as cold atomic dimers confined to a one-dimensional optical lattice with
sharp boundaries. 

\begin{table}[tb]
\caption{Coefficients of the effective range expansion for one-dimensional
dimer-wall scattering.}%
\label{1D_table}%
$%
\begin{tabular}
[c]{|c|c|c|c|}%
$m_{2}/m_{1}$ & $\kappa_{B}a_{R}$ & $\kappa_{B}r_{R}$ & $\kappa_{B}^{3}P_{R}%
$\\\hline\hline
1 & $0.4999(2)$ & $0.005(7)$ & $0.002(3)$\\
2 & $0.6065(2)$ & $-0.074(2)$ & $-0.006(2)$\\
4 & $0.8747(2)$ & $0.115(2)$ & $0.006(2)$\\
8 & $1.2149(2)$ & $0.460(2)$ & $0.008(2)$%
\end{tabular}
\ \ \ \ \ \ \ $\end{table}

\subsubsection{Numerical Results in Two Dimensions}

\begin{figure}[ht]
\centering
\includegraphics[width=0.8\textwidth]
{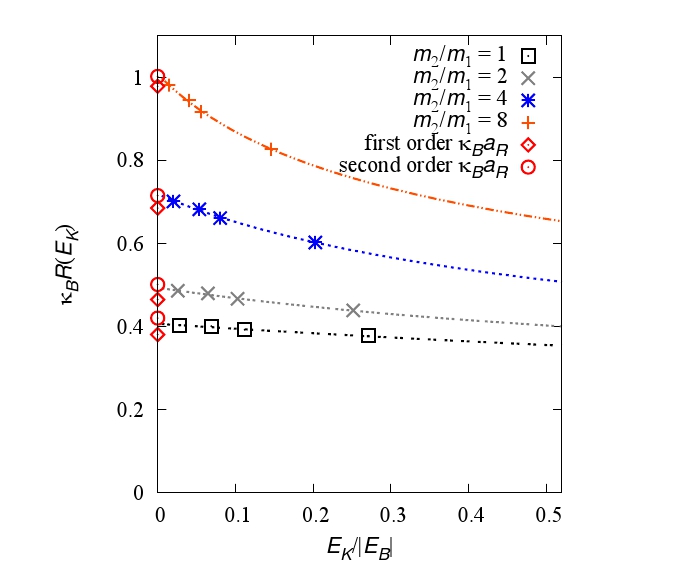}
\caption{Two-dimensional lattice results for the reflection radius vs dimer
kinetic energy in the shallow binding-energy limit. Also shown are first- and
second- order results for the expansion of $\kappa_{B}a_{R}$ described later
in the text.}
\label{latt_2D}
\end{figure}

Fig.~\ref{latt_2D} shows the reflection radius versus dimer kinetic energy for
the two-dimensional system in the shallow-binding limit. Again the results are
presented in dimensionless combinations, $\kappa_{B}R(E_{K})$ versus
$E_{K}/\left\vert E_{B}\right\vert $. The reflection radius is somewhat
smaller than in the one-dimensional case. This is reasonable considering the
difference between compression of a one-dimensional object versus compression
of a two-dimensional object along just one dimension. In the first case, we
expect the object to resist compression more since it does not have another
direction to expand along as it is compressed.\ As in the one-dimensional
case, we see the same dependence on the mass ratio.$\ $At large $m_{2}/m_{1}$
the reflection radius is large at small energies while becoming substantially
smaller with increasing energy.\ Fig.~\ref{latt_2D} also shows the first- and
second-order results for the asymptotic expansion of $\kappa_{B}a_{R}$ derived
from the general theory for shallow two-body bound state reflection.%

%TCIMACRO{\FRAME{ftbpFU}{2.2684in}{2.2174in}{0pt}{\Qcb{Two-dimensional lattice
%results for the reflection radius vs dimer kinetic energy in the shallow
%binding-energy limit. Also shown are first- and second- order results for the
%expansion of $\kappa_{B}a_{R}$ described later in the text.}}{\Qlb{latt_2D}%
%}{radius_vs_e_2d.eps}{\special{ language "Scientific Word";  type "GRAPHIC";
%display "USEDEF";  valid_file "F";  width 2.2684in;  height 2.2174in;
%depth 0pt;  original-width 5.5988in;  original-height 5.4993in;
%cropleft "0";  croptop "1";  cropright "1";  cropbottom "0";
%filename 'radius_vs_E_2D.eps';file-properties "XNPEU";}} }%
%BeginExpansion
%EndExpansion

The coefficients of the effective range expansion for the two-dimensional
system are given in Table~\ref{2D_table}. \ 

%

%TCIMACRO{\TeXButton{B}{\begin{table}[tbp] \centering}}%
%BeginExpansion
\begin{table}[h] \centering
%EndExpansion
\caption{Coefficients of the effective range expansion for two-dimensional dimer-wall scattering}\label{2D_table}%
$%
\begin{tabular}
[c]{|c|c|c|c|}%
$m_{2}/m_{1}$ & $\kappa_{B}a_{R}$ & $\kappa_{B}r_{R}$ & $\kappa_{B}^{3}P_{R}%
$\\\hline\hline
1 & $0.407(5)$ & $-0.10(28)$ & $0.00(11)$\\
2 & $0.494(5)$ & $-0.18(19)$ & $-0.01(7)$\\
4 & $0.718(6)$ & $-0.01(10)$ & $-0.00(3)$\\
8 & $1.009(9)$ & $0.29(6)$ & $-0.001(17)$%
\end{tabular}
\ \ $%

%TCIMACRO{\TeXButton{E}{\end{table}}}%
%BeginExpansion
\end{table}%
%EndExpansion

\subsubsection{Numerical Results in Three Dimensions}

\begin{figure}
\centering
\includegraphics[width=0.8\textwidth]{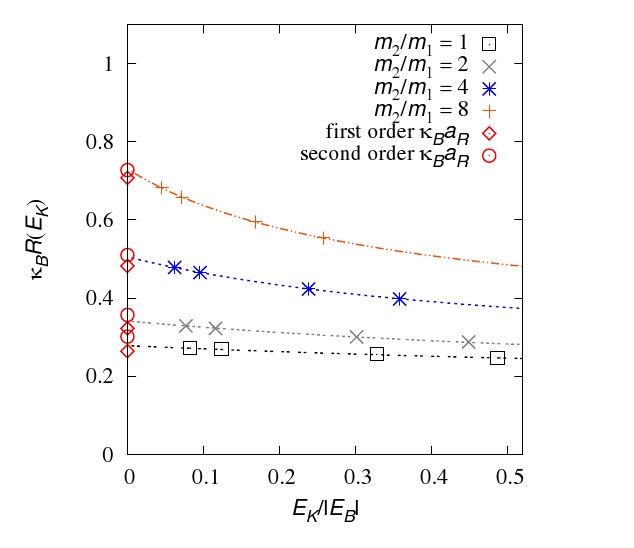}
\caption{Three-dimensional lattice results for the reflection radius versus
dimer kinetic energy in the shallow-binding limit. \ Also shown are first- and
second-order results for $\kappa_{B}a_{R}$ described in the text.}%
\label{latt_3D}
\end{figure}

Fig.~\ref{latt_3D} shows the reflection radius versus dimer kinetic energy for
the three-dimensional system in the shallow-binding limit. Again the results
are presented in dimensionless combinations, $\kappa_{B}R(E_{K})$ versus
$E_{K}/\left\vert E_{B}\right\vert $. The results are plotted at several
different mass ratios, $m_{2}/m_{1}$.\ Again the reflection radius is smaller
than in the two-dimensional case and nearly a factor of two smaller than in
the one-dimensional case.\ Just as in the two-dimensional case, this conforms
to our expectation that the three-dimensional object will resist compression
even less than the two dimensional object, since it has two unconstrained
dimensions it can expand along rather than just one.\ As in the one- and
two-dimensional cases, we once again find a direct relationship between mass
ratio and the derivative of the reflection radius with respect to the dimer
kinetic energy. Fig.~\ref{latt_3D} also shows the first- and second-order
results for the asymptotic expansion of $\kappa_{B}a_{R}$ derived from the
general theory for shallow two-body bound state reflection.%
%TCIMACRO{\FRAME{ftbpFU}{2.2675in}{2.2278in}{0pt}{\Qcb{Three-dimensional
%lattice results for the reflection radius versus dimer kinetic energy in the
%shallow-binding limit. \ Also shown are first- and second-order results for
%$\kappa_{B}a_{R}$ described later in the text.}}{\Qlb{latt_3D}}%
%{radius_vs_e_3d.eps}{\special{ language "Scientific Word";  type "GRAPHIC";
%width 2.2675in;  height 2.2278in;  depth 0pt;  original-width 5.5988in;
%original-height 5.4993in;  cropleft "0";  croptop "1";  cropright "1";
%filename 'HardwallPaper/hardwallPaperSept/radius_vs_E_3D.eps';file-properties "XNPEU";}%
%} }%
%BeginExpansion
%EndExpansion
The coefficients of the effective range expansion for the three-dimensional
system are given in Table~\ref{3D_table}. \ 

\begin{table}[h]
\caption{Coefficients of the effective range expansion for three-dimensional
dimer-wall scattering.}%
\label{3D_table}%
\begin{tabular}
[c]{|c|c|c|c|}%
$m_{2}/m_{1}$ & $\kappa_{B}a_{R}$ & $\kappa_{B}r_{R}$ & $\kappa_{B}^{3}P_{R}%
$\\\hline\hline
1 & $0.279(8)$ & $-0.38(9)$ & $-0.02(2)$\\
2 & $0.342(8)$ & $-0.44(6)$ & $-0.02(2)$\\
4 & $0.506(8)$ & $-0.27(4)$ & $-0.02(1)$\\
8 & $0.731(8)$ & $0.03(2)$ & $-0.005(5)$%
\end{tabular}
\ \ \ \ \ \ \end{table}

We now address what happens in the limit $m_{2}/m_{1}\rightarrow\infty
$.\ Consider the limit $m_{2}\rightarrow\infty$ with $m_{1}$ held fixed.\ In
this limit the effective potential converges to a non-vanishing finite-valued
function, while the mass of the dimer grows with $m_{2}$.\ One can check this
explicitly for the expressions in Eq.~(\ref{V1_1D}) and Eq.~(\ref{V1_3D}).
Given the exponential tail of the effective potential, the reflection radius
near threshold has a logarithmic dependence on $m_{2}/m_{1}$. We see this behavior in each of the plots in Fig. \ref{latt_1D}, \ref{latt_2D}, and \ref{latt_3D}. \ 
\subsubsection{\bigskip Alpha\ Particle}  

\begin{table}
\caption{Momenta and reflection radii for an alpha particle confined to a cube
of length $L$.}
\label{alpha}

\begin{tabular}
[c]{|c|c|c|}%
$L$ & $p(L)$ & $R[p(L)]$\\\hline\hline
$11.8$ fm & $81(9)$ MeV & $2.1(4)$ fm\\
$9.9$ fm & $97(10)$ MeV & $1.6(3)$ fm\\
$7.9$ fm & $118(10)$ MeV & $1.3(2)$ fm
\end{tabular}
\ \ \ \ \ \end{table}

%TCIMACRO{\FRAME{ftbpFU}{2.3912in}{2.4405in}{0pt}{\Qcb{Energy of the alpha
%particle confined to a cubic box of lengths 11.8 fm, 9.9 fm, and 7.9 fm.}%
%}{\Qlb{Alpha_Energy}}{helium4_wall.eps}{\special{ language "Scientific Word";
%type "GRAPHIC";  maintain-aspect-ratio TRUE;  display "USEDEF";
%valid_file "F";  width 2.3912in;  height 2.4405in;  depth 0pt;
%original-width 4.9in;  original-height 5.0004in;  cropleft "0";  croptop "1";
%cropright "1";  cropbottom "0";
%filename 'figuresForPaper/helium4_wall.eps';file-properties "XNPEU";}} }%
%BeginExpansion

\begin{figure}[ht]
\centering
\includegraphics[width=0.6\textwidth]{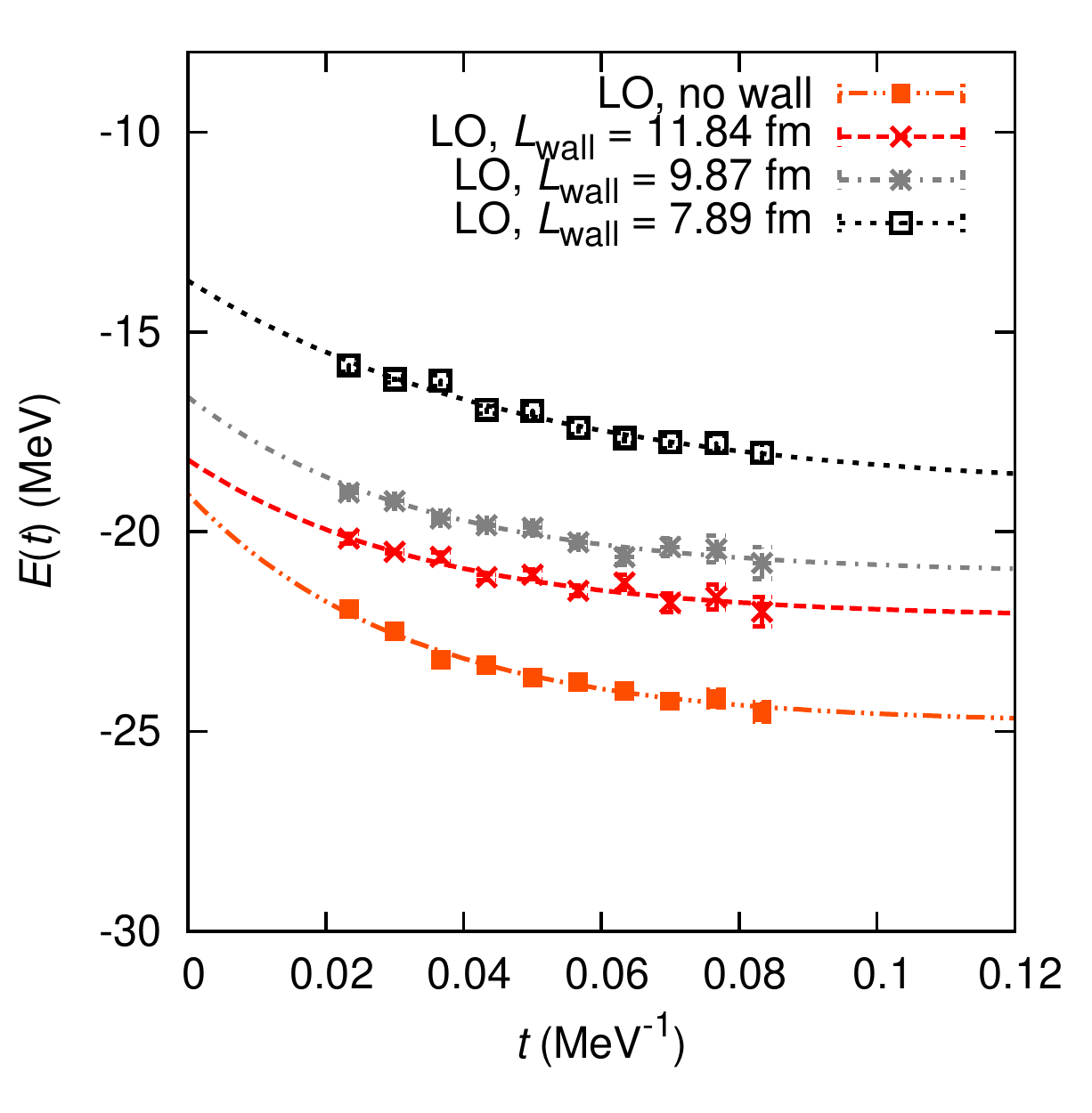}
\caption{Energy of an alpha particle confined to cubic box of lengths 11.8, 9.9, and 7.9 fm.}
\label{Alpha_Energy}
\end{figure}
%EndExpansion

Using lattice effective field theory, we have calculated the energy of an alpha particle in a cubic box at leading
order in chiral effective field theory at lattice spacing $a=1.97$ fm for cubic boxes of lengths $L=11.8$~fm,
$9.9$~fm, and $7.9$~fm.\ We use the same lattice action, algorithms, and codes
as in Ref.~\cite{Epelbaum:2011md}.\ A review of lattice effective field theory can be found in Ref.~\cite{Lee:2008fa}. The vanishing Dirichlet boundaries are
implemented as an essentially infinite potential energy at the wall
boundaries.\ We note that the ultraviolet divergences are independent of
long-distance boundary conditions, and so no new renormalization counterterms
are needed.\ In Fig. \ref{Alpha_Energy} we plot the energy expectation value of the alpha particle as a function of the Euclidean time projection. The exponential curves shown are best fits to a functional form 
\begin{equation}
E(t)=E_{0}+ce^{-\Delta Et}. \label{Euclidean_time}
\end{equation}
These capture the asymptotic behavior at large $t$.

Using the asymptotic form in Eq.~(\ref{Euclidean_time}) we extract the ground state energy $E_{0}$ for the given boundary conditions. Using Eqs.~(\ref{energy_length}) and (\ref{momentum_length}), we can calculate $P(L)$ and $R[P(L)]$. The results are shown in Table \ref{alpha}. The error bars in Table
\ref{alpha} are one standard deviation estimates which include both Monte
Carlo statistical errors and uncertainties due to extrapolation at large
Euclidean time. At leading order we find the root-mean-square (RMS) matter radius for point-like
nucleons to be $1.53(4)$~fm.\ Comparing the reflection radii in Table
\ref{alpha} to the RMS matter radius of the alpha particle,
we find that at low momenta the reflection radius of the alpha particle is
larger than the RMS matter radius. However for larger momenta we see a
substantial decrease in the alpha particle reflection
radius.\ 
This suggests that the alpha particle is quite compressible under confinement pressure.\ This appears consistent with
the observation that alpha clusters are compressed in size when confined within nuclei.\ Much 
more numerical work is planned to study alpha particles and other nuclei under confinement pressure.%

\section{Summary and Future Work}

In this paper we derived the phase shift due to the scattering of a two-particle bound state on a hard wall in the adiabatic limit and up to second order in an expansion of the effective Hamiltonian.  We have presented the effective Hamiltonian as an expansion in the parameters $e^{-\kappa_{B}r_{+}}$ and $e^{-\kappa_{B}r_{-}}$. We plotted the effective
potential in one, two, and three dimensions, and discussed the convergence of the expansion. For the equal mass case in 
one dimension we have presented an exact analytic solution using the Bethe Ansatz method. Additionaly, we presented
numerical and effective field theory calculations of the reflection radius in one, two, and three spatial dimensions. 
We then discussed the consistency among the analytic, numeric, and effective field theory calculations.
We saw that, as expected, the bound state becomes more readily compressible as the number of spatial dimensions 
increases. Furthermore, we noted an interesting mass ratio dependence of the reflection radius, noting that for larger 
mass ratios the reflection radius is larger but decreases faster as the kinetic energy of the bound state increases. 
Using lattice effective field theory we calculated the alpha particle energy in a cubic box for 
$L = 11.8$ fm, $9.9$ fm, and $7.9$ fm. We then calculated the alpha particle reflection radius and in our analysis
of the alpha particle compressibility we found that the alpha particle actually appears to compress rather easily.

We have discussed many aspects of the elastic scattering of quantum
bound states from a hard surface, paying particular attention to universal
behavior which is common to many different systems.\ There appear to be many
applications, ranging from experimental predictions for quantum dots and wells
to numerical calculations of nuclear structure and elastic deformation. The
theoretical techniques employed in this analysis may prove useful in
describing the effective field theory of other inhomogeneous systems. One
immediate extension of our three-dimensional analysis is to calculate the
deuteron reflection phase shift from \textit{ab initio} lattice chiral
effective field theory to verify the universal effective range expansion
coefficients in Table~\ref{3D_table}, as well as to measure the spin
dependence of the reflection phase shift resulting from the $D$-wave component
of the deuteron wavefunction. Some other interesting extensions include
investigating the inelastic threshold, and varying the boundary conditions
used. 

Another immediate extension is to verify the logarithmic $m_{2}/m_{1}$
threshold dependence in various physical systems. We expect this effect to be
prominent for halo nuclei with a heavy core and one satellite nucleon as well
as in quantum dots and wells for semiconductors with a large ratio between
hole and electron effective masses. It could also be reproduced experimentally
with heterogeneous cold atomic dimers consisting of one heavy and one light
alkali atom.

\textit{\ Financial support from the U.S. Department of Energy and NC State
GAANN Fellowship is acknowledged. \ Computational resources provided by the
J\"{u}lich Supercomputing Centre at Forschungszentrum J\"{u}lich.}

\bibliographystyle{apsrev}
\bibliography{References}

\appendix{}

\section{\label{Diagonal}Derivation of the Diagonal Adiabatic Correction}

We have already defined the diagonal adiabatic correction as%

\begin{equation}
T(X)=-\frac{1}{2M}\left\langle \psi_{X}\right\vert
\frac{\partial^{2}}{\partial X^{2}}\left\vert \psi_{X}\right\rangle,
\end{equation}
where%

\begin{equation}
\psi_{X}(\vec{r})=A_{\vec{r}_{-},\vec{r}_{+}}\phi
_{\vec{r}_{-},\vec{r}_{+}}(\vec{r}).
\end{equation}
The normalization condition gives%

\begin{align}
A_{\vec{r}_{-},\vec{r}_{+}}  &  =\left[  f_{d}\left(  \kappa_{\vec{r}_{-}%
,\vec{r}_{+}},\kappa_{\vec{r}_{-},\vec{r}_{+}},0\right)  -f_{d}\left(
\kappa_{\vec{r}_{-},\vec{r}_{+}},\kappa_{\vec{r}_{-},\vec{r}_{+}}%
,r_{-}\right)  -f_{d}\left(  \kappa_{\vec{r}_{-},\vec{r}_{+}},\kappa_{\vec
{r}_{-},\vec{r}_{+}},r_{+}\right)  \right. \nonumber\\
&  \left.  +2f_{d}\left(  \kappa_{\vec{r}_{-},\vec{r}_{+}},\kappa_{\vec{r}%
_{-},\vec{r}_{+}},r_{+}+r_{-}\right)  +\cdots\right]  ^{-1/2},%
\end{align}
where%

\begin{align}
\phi_{\vec{r}_{-},\vec{r}_{+}}(\vec{r})  &  =y_{d,0}(\kappa_{\vec{r}_{-}%
,\vec{r}_{+}},\left\vert \vec{r}\right\vert )-y_{d,0}(\kappa_{\vec{r}_{-}%
,\vec{r}_{+}},\left\vert \vec{r}-\vec{r}_{-}\right\vert )-y_{d,0}(\kappa
_{\vec{r}_{-},\vec{r}_{+}},\left\vert \vec{r}-\vec{r}_{+}\right\vert
)\nonumber\\
&  +y_{d,0}(\kappa_{\vec{r}_{-},\vec{r}_{+}},\left\vert \vec{r}-\vec{r}%
_{+}+\vec{r}_{-}\right\vert )+y_{d,0}(\kappa_{\vec{r}_{-},\vec{r}_{+}},\vec
{r}+\vec{r}_{+}-\vec{r}_{-})+\cdots
\end{align}
and $A_{\vec{r}_{-},\vec{r}_{+}}$ is the function that normalizes the relative
coordinate wavefunction, $\phi_{\vec{r}_{-},\vec{r}_{+}}(\vec{r}).$ The $d$-dimensional Yukawa function is%

\begin{equation}
y_{d,0}(\kappa,r)=\int\frac{d^{d}\vec{p}}{\left(  2\pi\right)  ^{d}}%
\frac{e^{-i\vec{p}\cdot\vec{r}}}{p^{2}+\kappa^{2}} \label{Yukawa_General},%
\end{equation}
and the overlap integral is%

\begin{equation}
f_{d}\left(  \kappa^{I},\kappa^{II},r\right)  =\int d^{d}\vec{r}^{\;\prime
}\;y_{d,0}(\kappa^{I},\left\vert \vec{r}^{\;\prime}\right\vert )y_{d,0}%
(\kappa^{II},\left\vert \vec{r}^{\;\prime}-\vec{r}\right\vert ),
\label{Overlap_General}%
\end{equation}
for two Yukawa functions whose centers are
separated by a spatial distance $r$.

We think of $\phi_{\vec{r}_{-},\vec{r}_{+}}(\vec{r})$ as a Yukawa function
centered at the origin, plus images from a single mirror reflection, plus
images from two mirror reflections, and so on. We expand $A_{\vec{r}_{-}%
,\vec{r}_{+}}$ by including more reflections at each increasing order. This
procedure gives%

\begin{equation}
A_{\vec{r}_{-},\vec{r}_{+}(0)}=\frac{1}{\left[  f_{d}\left(  \kappa
_{B},\kappa_{B},0\right)  \right]  ^{1/2}},%
\end{equation}
and%

\begin{align}
A_{\vec{r}_{-},\vec{r}_{+}(1)}  &  =\frac{1}{2}\frac{f_{d}\left(  \kappa
_{B},\kappa_{B},r_{-}\right)  +f_{d}\left(  \kappa_{B},\kappa_{B}%
,r_{+}\right)  }{\left[  f_{d}\left(  \kappa_{B},\kappa_{B},0\right)  \right]
^{3/2}}+\kappa_{\vec{r}_{-},\vec{r}_{+}(1)}\frac{\partial A_{\vec{r}_{-}%
,\vec{r}_{+}(0)}}{\partial\kappa_{B}}\nonumber\\
&  =\frac{1}{2}\frac{f_{d}\left(  \kappa_{B},\kappa_{B},r_{+}\right)
+f_{d}\left(  \kappa_{B},\kappa_{B},r_{-}\right)  -\kappa_{\vec{r}_{-},\vec
{r}_{+}(1)}\frac{\partial}{\partial\kappa_{B}}f_{d}\left(  \kappa_{B}%
,\kappa_{B},0\right)  }{\left[  f_{d}\left(  \kappa_{B},\kappa_{B},0\right)
\right]  ^{3/2}}.%
\end{align}
Since the relative coordinate wavefunction is normalized,%

\begin{align}
&  -\frac{1}{2M}\left\langle \psi_{X}\right\vert
\frac{\partial^{2}}{\partial X^{2}}\left\vert \psi_{X}\right\rangle \nonumber\\
&  =-\frac{\left(  \frac{\partial A_{\vec{r}_{-},\vec{r}_{+}}}{\partial
X}\right)  ^{2}}{2MA_{\vec{r}_{-},\vec{r}_{+}}^{2}}+\frac{A_{\vec{r}_{-}%
,\vec{r}_{+}}^{2}}{2M}\int_{x_{-}/2}^{x_{+}/2}dx_{d}\int d^{d-1}\vec{r}_{\bot
}\;\left[  \frac{\partial\phi_{\vec{r}_{-},\vec{r}_{+}}}{\partial X}(\vec
{r})\right]  ^{2}.%
\end{align}

We compute this term as an expansion in powers of $e^{-\kappa_{B}r_{+}}$ and
$e^{-\kappa_{B}r_{-}}$. At leading order we find%

\begin{align}
&  -\frac{1}{2M}\left\langle \psi_{X}\right\vert
\frac{\partial^{2}}{\partial X_{d}}\left\vert \psi_{X}\right\rangle _{(1)}\nonumber\\
&  =\frac{M}{2m_{1}^{2}f_{d}\left(  \kappa_{B},\kappa_{B},0\right)  }%
y_{d,0}\left(  \kappa_{B},r_{+}\right)  +\frac{M}{2m_{2}^{2}f_{d}\left(
\kappa_{B},\kappa_{B},0\right)  }y_{d,0}\left(  \kappa_{B},r_{-}\right)  .
\end{align}
At second order%

\begin{align}
&  -\frac{1}{2M}\left\langle \psi_{X}\right\vert
\frac{\partial^{2}}{\partial X^{2}}\left\vert \psi_{X}\right\rangle _{(2)}\nonumber\\
&  =T_{(A)}^{(2)}(X)+T_{(B)}^{(2)}(X)+T_{(C)}^{(2)}(X)\nonumber\\
&  +T_{(D1)}^{(2)}%
(X)+T_{(D2)}^{(2)}(X)+T_{(D3)}^{(2)}(X)+T_{(D4)}^{(2)}(X)+T_{(E)}%
^{(2)}(X),\label{TA_Deriv}%
\end{align}
where the first term in Eq.(\ref{TA_Deriv}) is%

\begin{equation}
T_{(A)}^{(2)}(X)=-\frac{f_{d}\left(  \kappa_{B},\kappa_{B},0\right)  }%
{2M}\left(  \frac{\partial A_{\vec{r}_{-},\vec{r}_{+}(1)}}{\partial
X}\right)  ^{2}+\cdots,
\end{equation}
and the second term%

\begin{align}
&  T_{(B)}^{(2)}(X)\nonumber\\
&  =-\frac{M}{\mu^{2}f_{d}\left(  \kappa_{B},\kappa_{B},0\right)  }%
y_{d,0}(\kappa_{B},r_{+}+r_{-})+\frac{r_{+}+r_{-}}{\mu f_{d}\left(  \kappa
_{B},\kappa_{B},0\right)  }y_{d,-1}(\kappa_{B},r_{+}+r_{-}),
\end{align}
contains terms due to two derivatives with respect to the spatial coordinate
without any contribution from the first-order corrections $A_{\vec{r}_{-}%
,\vec{r}_{+}(1)}$ and $\kappa_{\vec{r}_{-},\vec{r}_{+}(1)}$.%
\begin{align}
T_{(C)}^{(2)}(X) &  =\frac{M\kappa_{\vec{r}_{-},\vec{r}_{+}(1)}}{2m_{1}%
^{2}f_{d}\left(  \kappa,\kappa,0\right)  }\left.  \frac{\partial}%
{\partial\kappa}\left[  y_{d,0}\left(  \kappa,r_{+}\right)  \right]
\right\vert _{\kappa=\kappa_{B}}+\frac{M\kappa_{\vec{r}_{-},\vec{r}_{+}(1)}%
}{2m_{2}^{2}f_{d}\left(  \kappa,\kappa,0\right)  }\left.  \frac{\partial
}{\partial\kappa}\left[  y_{d,0}\left(  \kappa,r_{-}\right)  \right]
\right\vert _{\kappa=\kappa_{B}}\nonumber\\
&  +\frac{A_{\vec{r}_{-},\vec{r}_{+}(1)}}{\left[  f_{d}\left(  \kappa
_{B},\kappa_{B},0\right)  \right]  ^{1/2}}\left[  \frac{M}{m_{1}^{2}}%
y_{d,0}\left(  \kappa_{B},r_{+}\right)  +\frac{M}{m_{2}^{2}}y_{d,0}\left(
\kappa_{B},r_{-}\right)  \right]  ,
\end{align}
contains terms proportional to $\kappa_{\vec{r}_{-},\vec{r}_{+}(1)}$ and
$A_{\vec{r}_{-},\vec{r}_{+}(1)}$.%

\begin{align}
&  T_{(D1)}^{(2)}(X)\nonumber\\
&  =-\frac{\kappa_{B}}{m_{1}f_{d}\left(  \kappa_{B},\kappa_{B},0\right)
}\frac{\partial\kappa_{\vec{r}_{-},\vec{r}_{+}(1)}}{\partial X}\left[
\frac{r_{+}^{2}}{8}y_{d,1}\left(  \kappa_{B},r_{+}\right)  +\frac{r_{+}}%
{2}y_{d,2}\left(  \kappa_{B},r_{+}\right)  +\frac{3}{4}y_{d,3}\left(
\kappa_{B},r_{+}\right)  \right]  ,\label{TD1_2}%
\end{align}%
\begin{align}
&  T_{(D2)}^{(2)}(X)\nonumber\\
&  =-\frac{\kappa_{B}}{m_{2}f_{d}\left(  \kappa_{B},\kappa_{B},0\right)
}\frac{\partial\kappa_{\vec{r}_{-},\vec{r}_{+}(1)}}{\partial X}\left[
\frac{r_{-}^{2}}{8}y_{d,1}\left(  \kappa_{B},r_{-}\right)  +\frac{r_{-}}%
{2}y_{d,2}\left(  \kappa_{B},r_{-}\right)  +\frac{3}{4}y_{d,3}\left(
\kappa_{B},r_{-}\right)  \right]  ,
\end{align}%
\begin{equation}
T_{(D3)}^{(2)}(X)=\frac{\kappa_{B}}{4m_{1}f_{d}\left(  \kappa_{B},\kappa
_{B},0\right)  }\frac{\partial\kappa_{\vec{r}_{-},\vec{r}_{+}(1)}}{\partial
X}\left[  r_{+}y_{d,2}\left(  \kappa_{B},r_{+}\right)  +3y_{d,3}\left(
\kappa_{B},r_{+}\right)  \right]  ,
\end{equation}

and%
\begin{equation}
T_{(D4)}^{(2)}(X)=\frac{\kappa_{B}}{4m_{2}f_{d}\left(  \kappa_{B},\kappa
_{B},0\right)  }\frac{\partial\kappa_{\vec{r}_{-},\vec{r}_{+}(1)}}{\partial
X}\left[  r_{-}y_{d,2}\left(  \kappa,r_{-}\right)  +3y_{d,3}\left(
\kappa,r_{-}\right)  \right],  
\end{equation}

contain terms due to a derivative with respect to $\kappa$ as well as a
spatial derivative. Finally,

\begin{align}
&  T_{(E)}^{(2)}(X)\nonumber\\
&  =\frac{\left(  \frac{\partial\kappa_{\vec{r}_{-},\vec{r}_{+}(1)}}{\partial
X}\right)  ^{2}}{24M\kappa_{B}^{3}f_{d}\left(  \kappa_{B},\kappa_{B},0\right)
}\left[  -3\frac{\partial y_{d,0}}{\partial\kappa_{B}}(\kappa_{B}%
,0)+3\kappa\frac{\partial^{2}y_{d,0}}{\partial\kappa_{B}^{2}}(\kappa
_{B},0)-\kappa^{2}\frac{\partial^{3}y_{d,0}}{\partial\kappa_{B}^{3}}%
(\kappa_{B},0)\right]  ,
\end{align}

contains terms due to two derivatives with respect to $\kappa$.

\section{\label{Diagonal1D}Diagonal Adiabatic Correction in One
Dimension}

\bigskip Using Eq.~(\ref{Yukawa_General}) and Eq.~(\ref{Overlap_General}) we
compute the adiabatic correction in one spatial dimension. The first term in
Eq.~(\ref{TA_Deriv})%

\begin{align}
&  T_{(A)}^{(2)}(X)=\nonumber\\
- &  \frac{e^{-2\kappa_{B}r_{+}}\kappa_{B}^{2}\left(  m_{1}+m_{2}\right)
\left[  e^{\kappa_{B}(r_{+}+r_{-})}m_{1}\left(  3+\kappa_{B}r_{-}\right)
+m_{2}\left(  3-\kappa_{B}r_{+}\right)  \right]  ^{2}}{2m_{1}^{2}m_{2}^{2}},
\end{align}
and the second term in Eq.(\ref{TA_Deriv}), due to two spatial derivatives but not
including contributions due to $A_{\vec{r}_{-},\vec{r}_{+}(1)}$ and
$\kappa_{\vec{r}_{-},\vec{r}_{+}(1)}$, is%

\begin{align}
&  T_{(B)}^{(2)}(X)=\nonumber\\
&  \frac{2e^{-\kappa_{B}(r_{+}+r_{-})}\kappa_{B}^{2}\left(  m_{1}%
+m_{2}\right)  \left(  -m_{1}^{2}-m_{2}^{2}+m_{1}m_{2}\left[  \kappa_{B}%
r_{+}+\kappa_{B}r_{-}-2\right]  \right)  }{m_{1}^{2}m_{2}^{2}}.
\end{align}
The term proportional to $\kappa_{\vec{r}_{-},\vec{r}_{+}(1)}$ and
$A_{\vec{r}_{-},\vec{r}_{+}(1)}$ is%

\begin{align}
&  T_{(C)}^{(2)}(X)\nonumber\\
&  =\frac{e^{-2\kappa_{B}r_{-}}}{m_{2}^{2}}\kappa_{B}^{2}\left(  m_{1}%
+m_{2}\right)  \left(  2\kappa_{B}r_{-}-1\right)  +\frac{e^{-2\kappa_{B}r_{+}%
}}{m_{1}^{2}}\kappa_{B}^{2}\left(  m_{1}+m_{2}\right)  \left(  2\kappa
_{B}r_{+}-1\right) \nonumber\\
&  +\frac{e^{-\kappa_{B}(r_{+}+r_{-})}\left(  m_{1}^{2}+m_{2}^{2}\right)
}{m_{1}^{2}m_{2}^{2}}\kappa_{B}^{2}\left(  m_{1}+m_{2}\right)  \left(
2\kappa_{B}r_{+}+2\kappa_{B}r_{-}-1\right).
\end{align}
The four terms due to one derivative with respect to $\kappa$ and one spatial
derivative are%

\begin{align}
&  T_{(D1)}^{(2)}(X)=\\
&  -\frac{\kappa_{B}^{2}\left(  m_{1}+m_{2}\right)  \left(  6+4\kappa_{B}%
r_{+}+\kappa_{B}^{2}r_{+}^{2}\right)  }{2m_{1}^{2}m_{2}^{2}}\left(
m_{2}e^{-2\kappa_{B}r_{+}}+m_{1}e^{-\kappa_{B}(r_{+}+r_{-})}\right)
,\nonumber
\end{align}

\begin{align}
&  T_{(D2)}^{(2)}(X)=\nonumber\\
&  -\frac{\kappa_{B}^{2}\left(  m_{1}+m_{2}\right)  \left(  9-4\kappa_{B}%
r_{+}+\kappa_{B}^{2}r_{-}^{2}\right)  }{2m_{1}m_{2}^{2}}\left(  m_{2}%
e^{-\kappa_{B}(r_{+}+r_{-})}+m_{1}e^{-2\kappa_{B}r_{-}}\right)  ,
\end{align}

\begin{equation}
T_{(D3)}^{(2)}(X)=\frac{\kappa_{B}^{2}\left(  m_{1}+m_{2}\right)  \left(
3+\kappa_{B}r_{+}\right)  }{m_{1}^{2}m_{2}}\left(  m_{2}e^{-2\kappa_{B}r_{+}%
}+m_{1}e^{-\kappa_{B}(r_{+}+r_{-})}\right)  ,
\end{equation}
and%

\begin{equation}
T_{(D4)}^{(2)}(X)=\frac{\kappa_{B}^{2}\left(  m_{1}+m_{2}\right)  \left(
3+\kappa_{B}r_{-}\right)  }{m_{1}m_{2}^{2}}\left(  m_{2}e^{-\kappa_{B}%
(r_{+}+r_{-})}+m_{1}e^{-2\kappa_{B}r_{-}}\right)  .
\end{equation}
The term containing contributions due to two derivatives with respect to $\kappa$ is%

\begin{equation}
T_{(E)}^{(2)}(X)=5M\kappa_{B}^{2}\left(  \frac{e^{-2\kappa_{B}r_{+}}}%
{m_{1}^{2}}+\frac{e^{-2\kappa_{B}r_{-}}}{m_{2}^{2}}\right)  +\frac
{10e^{-\kappa_{B}(r_{+}+r_{-})}\kappa_{B}^{2}}{\mu}.
\end{equation}

\section{\label{Diagonal2D}Diagonal Adiabatic Correction in Two Dimensions}

Using Eq.~(\ref{Yukawa_General}) and Eq.~(\ref{Overlap_General}) we compute the
adiabatic correction in two spatial dimensions. We find that in two dimensions
the first order correction to the binding momentum is%

\begin{equation}
\kappa_{\vec{r}_{-},\vec{r}_{+}(1)}=-\kappa_{B}\left[  K_{0}(\kappa_{B}%
r_{+})+K_{0}(\kappa_{B}r_{-})\right]  ,
\end{equation}
the second order correction to the binding momentum is%
\begin{align}
\kappa_{\vec{r}_{-},\vec{r}_{+}(2)} &  =\frac{\kappa_{B}}{2}\left\{
K_{0}(\kappa_{B}r_{+})^{2}+K_{0}(\kappa_{B}r_{-})^{2}+4K_{0}\left[  \kappa
_{B}\left(  r_{+}+r_{-}\right)  \right]  \right\}  \nonumber\\
&  -\kappa_{B}K_{0}(\kappa_{B}r_{+})\left[  \kappa_{B}r_{+}K_{1}(\kappa
_{B}r_{+})+\kappa_{B}r_{-}K_{1}(\kappa_{B}r_{-})-K_{0}(\kappa_{B}%
r_{-})\right]  \nonumber\\
&  -\kappa_{B}^{2}K_{0}(\kappa_{B}r_{-})\left[  r_{+}K_{1}(\kappa_{B}%
r_{+})+\kappa_{B}r_{-}K_{1}(\kappa_{B}r_{-})\right]  ,
\end{align}
and the generalized Yukawa function in two dimensions is%
\begin{equation}
y_{2,0}(\kappa,r)=\frac{1}{2\pi}K_{0}(\kappa r).
\end{equation}
The first term in Eq.~(\ref{TA_Deriv})%

\begin{align}
&  T_{(A)}^{(2)}(X)\nonumber\\
&  =-\frac{M\kappa_{B}^{2}}{8m_{1}^{2}m_{2}^{2}}\left\{  \kappa_{B}m_{1}%
r_{-}K_{0}(\kappa_{B}r_{-})+\kappa_{B}m_{2}r_{+}K_{0}(\kappa_{B}r_{+}%
)-6m_{1}K_{1}(\kappa_{B}r_{-})\right. \nonumber\\
&  \left.  -6m_{2}K_{1}(\kappa_{B}r_{+})+\kappa_{B}m_{1}r_{-}K_{2}(\kappa
_{B}r_{-})+\kappa_{B}m_{2}r_{+}K_{2}(\kappa_{B}r_{+})\right\}  ^{2},
\end{align}
and the second term due to two spatial derivatives but not including
contributions due to $A_{\vec{r}_{-},\vec{r}_{+}(1)}$ and $\kappa_{\vec
{r}_{-},\vec{r}_{+}(1)}$, is%

\begin{align}
&  T_{(B)}^{(2)}(X)\nonumber\\
&  =-\frac{2M\kappa_{B}^{2}}{m_{1}^{2}m_{2}^{2}}\left\{  -(m_{1}+m_{2}%
)^{2}K_{0}[\kappa_{B}(r_{+}+r_{-})]\right. \nonumber\\
&  \left.  +\kappa_{B}m_{1}m_{2}(r_{+}+r_{-})K_{1}[\kappa_{B}(r_{+}%
+r_{-})]\right\}.
\end{align}
The term proportional to $\kappa_{\vec{r}_{-},\vec{r}_{+}(1)}$ and
$A_{\vec{r}_{-},\vec{r}_{+}(1)}$ is%

\begin{align}
T_{(C)}^{(2)}(X)  &  =-\frac{\kappa^{3}M\left[  K_{0}(\kappa r_{+}%
)+K_{0}(\kappa r_{-})\right]  }{m_{1}^{2}}\left.  \frac{\partial}%
{\partial\kappa}\left[  K_{0}(\kappa r_{+})\right]  \right\vert _{\kappa
=\kappa_{B}}\nonumber\\
&  +-\frac{\kappa^{3}M\left[  K_{0}(\kappa r_{+})+K_{0}(\kappa r_{-})\right]
}{m_{2}^{2}}\left.  \frac{\partial}{\partial\kappa}\left[  K_{0}(\kappa
r_{-})\right]  \right\vert _{\kappa=\kappa_{B}}\nonumber\\
&  +2\kappa\sqrt{\pi}A_{\vec{r}_{-},\vec{r}_{+}(1)}\left[  \frac{M}{2\pi
m_{1}^{2}}K_{0}(\kappa_{B}r_{+})+\frac{M}{2\pi m_{2}^{2}}K_{0}(\kappa_{B}%
r_{-})\right]  ,
\end{align}
where%

\begin{equation}
A_{\vec{r}_{-},\vec{r}_{+}(1)}=\sqrt{\pi}\kappa_{B}\left[  -2K_{0}(\kappa_{B}r_{-})-2K_{0}(\kappa_{B}%
r_{+})+\kappa_{B}r_{-}K_{1}(\kappa_{B}r_{-})+\kappa_{B}r_{+}K_{1}(\kappa
_{B}r_{+})\right]  ^{-1/2}.
\end{equation}
The four terms due to one derivative with respect to $\kappa$ and one spatial
derivative are%

\begin{align}
&  T_{(D1)}^{(2)}(X)\nonumber\\
&  =-\frac{M\kappa_{B}^{2}}{4m_{1}^{2}m_{2}}\left[  m_{1}K_{1}(\kappa_{B}%
r_{-})+m_{2}K_{1}(\kappa_{B}r_{+})\right]  \left\{  -3\kappa_{B}^{2}r_{+}%
^{2}G_{1,3}^{2,1}\left[  \frac{\kappa_{B}r_{+}}{2},\frac{1}{2}\left\vert
\begin{array}
[c]{c}%
0\\
-\frac{1}{2},\frac{1}{2},-1
\end{array}
\right.  \right]  \right. \nonumber\\
&  \left.  +3\pi\left[  \kappa_{B}r_{+}K_{0}(\kappa_{B}r_{+})+2K_{1}%
(\kappa_{B}r_{+})\right]  L_{1}(\kappa_{B}r_{+})+\kappa_{B}r_{+}K_{1}%
(\kappa_{B}r_{+})\left[  4\kappa_{B}r_{+}+3\pi L_{2}(\kappa_{B}r_{+})\right]
\right\}  ,
\end{align}
where $L_{\alpha}$ is the modified Stuve function and $G_{p,q}^{m,n}$ is the
generalized Meijer G-function.%

\begin{align}
&  T_{(D2)}^{(2)}(X)\nonumber\\
&  =-\frac{M\kappa_{B}^{2}}{4m_{1}m_{2}^{2}}\left[  m_{1}K_{1}(\kappa_{B}%
r_{-})+m_{2}K_{1}(\kappa_{B}r_{+})\right]  \left\{  -3\kappa_{B}^{2}r_{-}%
^{2}G_{1,3}^{2,1}\left[  \frac{\kappa_{B}r_{-}}{2},\frac{1}{2}\left\vert
\begin{array}
[c]{c}%
0\\
-\frac{1}{2},\frac{1}{2},-1
\end{array}
\right.  \right]  \right. \nonumber\\
&  \left.  +3\pi\left[  \kappa_{B}r_{-}K_{0}(\kappa_{B}r_{-})+2K_{1}%
(\kappa_{B}r_{-})\right]  L_{1}(\kappa_{B}r_{-})+\kappa_{B}r_{-}K_{1}%
(\kappa_{B}r_{-})\left[  4\kappa_{B}r_{-}+3\pi L_{2}(\kappa_{B}r_{-})\right]
\right\}  .
\end{align}

\begin{equation}
T_{(D3)}^{(2)}(X)=\frac{\pi\kappa_{B}^{2}}{m_{1}}\frac{\partial\kappa_{\vec
{r}_{-},\vec{r}_{+}(1)}}{\partial X}\left[  r_{+}y_{2,2}\left(  \kappa
_{B},r_{+}\right)  +3y_{2,3}\left(  \kappa_{B},r_{+}\right)  \right]  ,
\end{equation}
where the spatial derivative of the first order correction to the binding
momentum is%
\begin{equation}
\frac{\partial\kappa_{\vec{r}_{-},\vec{r}_{+}(1)}}{\partial X}=\frac
{2M}{m_{1}}\kappa_{B}^{2}K_{1}(\kappa_{B}r_{+})+\frac{2M}{m_{2}}\kappa_{B}%
^{2}K_{1}(\kappa_{B}r_{-}).
\end{equation}
The generalized Yukawa function $y_{2,2}(\kappa,r)$ is%

\begin{equation}
y_{2,2}(\kappa,r)=-\frac{r^{2}}{4}\left[  \frac{1}{\kappa r}-K_{0}(\kappa
r)L_{-1}(\kappa r)-K_{1}(\kappa r)L_{0}(\kappa r)\right]  +\frac{r}{2\pi
\kappa}K_{1}(\kappa r),
\end{equation}
the generalized Yukawa function $y_{2,3}(\kappa,r)$ is%
\begin{align}
y_{2,3}(\kappa,r)  &  =\frac{r^{3}}{8}\left[  \frac{1}{\kappa r}-K_{0}(\kappa
r)L_{-1}(\kappa r)-K_{1}(\kappa r)L_{0}(\kappa r)\right] \nonumber\\
&  +\frac{r}{8\kappa^{2}}K_{0}(\kappa r)L_{1}(\kappa r)\nonumber\\
&  +\frac{1}{24\pi\kappa^{3}}K_{1}(\kappa r)\left[  -4\kappa^{2}r^{2}+6\pi
L_{1}(\kappa r)+3\pi\kappa rL_{2}(\kappa r)\right] \nonumber\\
&  -\frac{r^{2}}{8\pi\kappa}G_{1,3}^{2,1}\left[  \frac{\kappa r}{2},\frac
{1}{2}\left\vert
\begin{array}
[c]{c}%
0\\
-\frac{1}{2},\frac{1}{2},-1
\end{array}
\right.  \right]  ,
\end{align}

and%

\begin{equation}
T_{(D4)}^{(2)}(X)=\frac{\pi\kappa_{B}^{2}}{m_{2}}\frac{\partial\kappa_{\vec
{r}_{-},\vec{r}_{+}(1)}}{\partial X}\left[  r_{-}y_{2,2}\left(  \kappa
,r_{-}\right)  +3y_{2,3}\left(  \kappa,r_{-}\right)  \right]  .
\end{equation}

The term containing contributions due to two derivatives with respect to $\kappa$ is%

\begin{equation}
T_{(E)}^{(2)}(X)=\frac{8\kappa_{B}^{2}\left(  m_{1}+m_{2}\right)  \left[
m_{1}K_{1}(\kappa_{B}r_{-})+m_{2}K_{1}(\kappa_{B}r_{+})\right]  ^{2}}%
{3m_{1}^{2}m_{2}^{2}}.
\end{equation}

\section{\label{Diagonal3D}Diagonal Adiabatic Correction in Three Dimensions}

For the effective Hamiltonian in three dimensions $\kappa_{\vec{r}_{-},\vec
{r}_{+}(1)},$ $\kappa_{\vec{r}_{-},\vec{r}_{+}(2)},$ $T_{(A)}^{(2)}(X),$
$T_{(B)}^{(2)}(X),$ $T_{(C)}^{(2)}(X),$ $T_{(D1-D4)}^{(2)}(X),$ and $T_{(E)}^{(2)}(X)$
are defined as follows. The first order correction to the binding momentum is%

\begin{equation}
\kappa_{\vec{r}_{-},\vec{r}_{+}(1)}=\frac{y_{3,0}\left(  \kappa_{B}%
,r_{+}\right)  +y_{3,0}\left(  \kappa_{B},r_{-}\right)  }{\frac{\partial
}{\partial\kappa_{B}}y_{3,0}\left(  \kappa_{B},0\right)  },
\end{equation}
the second order correction to the binding momentum is%

\begin{align}
&  \kappa_{\vec{r}_{-},\vec{r}_{+}(2)}\nonumber\\
&  =\frac{1}{\frac{\partial}{\partial\kappa_{B}}y_{3,0}\left(  \kappa
_{B},0\right)  }\left\{  \frac{y_{3,0}\left(  \kappa_{B},r_{+}\right)
+y_{3,0}\left(  \kappa_{B},r_{-}\right)  }{\frac{\partial}{\partial\kappa_{B}%
}y_{3,0}\left(  \kappa_{B},0\right)  }\frac{\partial}{\partial\kappa_{B}%
}\left[  y_{3,0}\left(  \kappa_{B},r_{+}\right)  +y_{3,0}\left(  \kappa
_{B},r_{-}\right)  \right]  \right. \nonumber\\
&  \left.  -2y_{3,0}\left(  \kappa_{B},r_{+}+r_{-}\right)  -\frac{\left[
y_{3,0}\left(  \kappa_{B},r_{+}\right)  +y_{3,0}\left(  \kappa_{B}%
,r_{-}\right)  \right]  ^{2}}{2\left[  \frac{\partial}{\partial\kappa_{B}%
}y_{3,0}\left(  \kappa_{B},0\right)  \right]  ^{2}}\frac{\partial^{2}%
}{\partial\kappa_{B}^{2}}y_{3,0}\left(  \kappa_{B},0\right)  \right\}  ,
\end{align}
and the generalized Yukawa function in three spatial dimensions is%
\begin{equation}
y_{3,0}\left(  \kappa,r\right)  =\frac{e^{-\kappa r}}{4\pi r}.
\end{equation}
The first term in Eq.~(\ref{TA_Deriv}) is%

\begin{align}
&  T_{(A)}^{(2)}(X)\nonumber\\
&  =-\frac{f_{3}\left(  \kappa_{B},\kappa_{B},0\right)  }{2M}\left(
\frac{\partial A_{\vec{r}_{-},\vec{r}_{+}(1)}}{\partial X}\right)  ^{2},
\end{align}
where the overlap integral in three spatial dimensions is%

\begin{equation}
f_{3}\left(  \kappa,\kappa,r\right)  =\frac{e^{-\kappa r}}{8\pi\kappa},
\end{equation}
and the spatial derivative of the first order correction to the wavefunction
normalization is%
\begin{align}
&  \frac{\partial A_{\vec{r}_{-},\vec{r}_{+}(1)}}{\partial X}\nonumber\\
&  =\frac{1}{2}\frac{\frac{2M}{m_{1}}\frac{\partial}{\partial r_{+}}%
f_{3}\left(  \kappa_{B},\kappa_{B},r_{+}\right)  +\frac{2M}{m_{2}}%
\frac{\partial}{\partial r_{-}}f_{3}\left(  \kappa_{B},\kappa_{B}%
,r_{-}\right)  -\frac{\partial\kappa_{(1)}}{\partial X}\frac{\partial
}{\partial\kappa_{B}}f_{3}\left(  \kappa_{B},\kappa_{B},0\right)  }{\left[
f_{3}\left(  \kappa_{B},\kappa_{B},0\right)  \right]  ^{3/2}}.
\end{align}
The next term, with contributions due solely to two spatial derivatives, is

\begin{align}
&  T_{(B)}^{(2)}(X)\nonumber\\
&  =-\frac{M}{\mu^{2}f_{3}\left(  \kappa_{B},\kappa_{B},0\right)  }%
y_{3,0}(\kappa_{B},r_{+}+r_{-})+\frac{r_{+}+r_{-}}{\mu f_{3}\left(  \kappa
_{B},\kappa_{B},0\right)  }y_{3,-1}(\kappa_{B},r_{+}+r_{-}), \label{TB_2}%
\end{align}
where the generalized Yukawa function, $y_{3,-1}(\kappa,r)$, is%

\begin{equation}
y_{3,-1}(\kappa,r)=-\frac{\partial}{\partial\kappa_{B}}y_{3,0}(\kappa_{B},r).
\end{equation}
The term containing contributions due to $\kappa_{\vec{r}_{-},\vec{r}_{+}%
(1)}$ and $A_{\vec{r}_{-},\vec{r}_{+}(1)}$ is%

\begin{align}
&  T_{(C)}^{(2)}(X)\nonumber\\
&  =\frac{M\kappa_{\vec{r}_{-},\vec{r}_{+}(1)}}{2m_{1}^{2}f_{3}\left(
\kappa,\kappa,0\right)  }\left.  \frac{\partial}{\partial\kappa}\left[
y_{3,0}\left(  \kappa,r_{+}\right)  \right]  \right\vert _{\kappa=\kappa_{B}%
}\nonumber\\
&  +\frac{M\kappa_{\vec{r}_{-},\vec{r}_{+}(1)}}{2m_{2}^{2}f_{3}\left(
\kappa,\kappa,0\right)  }\left.  \frac{\partial}{\partial\kappa}\left[
y_{3,0}\left(  \kappa,r_{-}\right)  \right]  \right\vert _{\kappa=\kappa_{B}%
}\nonumber\\
&  +\frac{A_{\vec{r}_{-},\vec{r}_{+}(1)}}{\left[  f_{3}\left(  \kappa
_{B},\kappa_{B},0\right)  \right]  ^{1/2}}\left[  \frac{M}{m_{1}^{2}}%
y_{3,0}\left(  \kappa_{B},r_{+}\right)  +\frac{M}{m_{2}^{2}}y_{3,0}\left(
\kappa_{B},r_{-}\right)  \right]  .
\end{align}
The four terms that result from taking one derivative with respect to $\kappa$
and one spatial derivative are%

\begin{align}
&  T_{(D1)}^{(2)}(X)\nonumber\\
&  =-\frac{\kappa_{B}}{m_{1}f_{3}\left(  \kappa_{B},\kappa_{B},0\right)
}\frac{\partial\kappa_{\vec{r}_{-},\vec{r}_{+}(1)}}{\partial X}\left[
\frac{r_{+}^{2}}{8}y_{3,1}\left(  \kappa_{B},r_{+}\right)  +\frac{r_{+}}%
{2}y_{3,2}\left(  \kappa_{B},r_{+}\right)  +\frac{3}{4}y_{3,3}\left(
\kappa_{B},r_{+}\right)  \right]  ,
\end{align}
where the spatial derivative of the first order correction to $\kappa_{\vec
{r}_{-},\vec{r}_{+}}$ is%

\begin{equation}
\frac{\partial\kappa_{\vec{r}_{-},\vec{r}_{+}(1)}}{\partial X}=\frac
{2M\left[  \frac{1}{m_{1}}\frac{\partial}{\partial r_{+}}y_{3,0}\left(
\kappa_{B},r_{+}\right)  +\frac{1}{m_{2}}\frac{\partial}{\partial r_{+}%
}y_{3,0}\left(  \kappa_{B},r_{-}\right)  \right]  }{\frac{\partial}%
{\partial\kappa_{B}}y_{3,0}\left(  \kappa_{B},0\right)  },
\end{equation}
and the generalized Yukawa functions, $y_{3,1}(\kappa,r)$, $y_{3,2}(\kappa
,r)$, and $y_{3,3}(\kappa,r)$, are%

\begin{equation}
y_{3,1}(\kappa,r)=-\frac{1}{4\pi}\operatorname{Ei}(-\kappa r),
\end{equation}%
\begin{equation}
y_{3,2}(\kappa,r)=\frac{e^{-\kappa r}}{4\pi\kappa}+\frac{r}{4\pi
}\operatorname{Ei}(-\kappa r),
\end{equation}
and%
\begin{equation}
y_{3,3}(\kappa,r)=\frac{e^{-\kappa r}}{8\pi\kappa^{2}}\left(  1-\kappa
r\right)  -\frac{r^{2}}{8\pi}\operatorname{Ei}(-\kappa r).
\end{equation}
Here $\operatorname{Ei}$ is the exponential integral function.%

\begin{align}
&  T_{(D2)}^{(2)}(X)\nonumber\\
&  =-\frac{\kappa_{B}}{m_{2}f_{3}\left(  \kappa_{B},\kappa_{B},0\right)
}\frac{\partial\kappa_{\vec{r}_{-},\vec{r}_{+}(1)}}{\partial X}\left[
\frac{r_{-}^{2}}{8}y_{3,1}\left(  \kappa_{B},r_{-}\right)  +\frac{r_{-}}%
{2}y_{3,2}\left(  \kappa_{B},r_{-}\right)  +\frac{3}{4}y_{3,3}\left(
\kappa_{B},r_{-}\right)  \right]  ,
\end{align}%
\begin{equation}
T_{(D3)}^{(2)}(X)=\frac{\kappa_{B}}{4m_{1}f_{3}\left(  \kappa_{B},\kappa
_{B},0\right)  }\frac{\partial\kappa_{\vec{r}_{-},\vec{r}_{+}(1)}}{\partial
X}\left[  r_{+}y_{3,2}\left(  \kappa_{B},r_{+}\right)  +3y_{3,3}\left(
\kappa_{B},r_{+}\right)  \right]  ,
\end{equation}
and%
\begin{equation}
T_{(D4)}^{(2)}(X)=\frac{\kappa_{B}}{4m_{2}f_{3}\left(  \kappa_{B},\kappa
_{B},0\right)  }\frac{\partial\kappa_{\vec{r}_{-},\vec{r}_{+}(1)}}{\partial
X}\left[  r_{-}y_{3,2}\left(  \kappa,r_{-}\right)  +3y_{3,3}\left(
\kappa,r_{-}\right)  \right]  .
\end{equation}
The term containing contributions due to two derivatives with respect to $\kappa$ is%

\begin{equation}
T_{(E)}^{(2)}(X)=\frac{\left(  \frac{\partial\kappa_{\vec{r}_{-},\vec{r}%
_{+}(1)}}{\partial X}\right)  ^{2}}{24M\kappa_{B}^{3}f_{3}\left(  \kappa
_{B},\kappa_{B},0\right)  }\left[  -3\frac{\partial y_{3,0}}{\partial
\kappa_{B}}(\kappa_{B},0)+3\kappa\frac{\partial^{2}y_{3,0}}{\partial\kappa
_{B}^{2}}(\kappa_{B},0)-\kappa^{2}\frac{\partial^{3}y_{3,0}}{\partial
\kappa_{B}^{3}}(\kappa_{B},0)\right]  .
\end{equation}

\section{\label{Extrap}Summary of Continuum Extrapolations}
In one dimension for mass ratio $m_{2}/m_{1}=1$ we consider $L_{z}|c|=8,10,20,$ 
and for $L_{z}|c|=10$ we consider the ground state as well as the first and second excited states.
Table \ref{1D_Extrap_ratio_1} shows a summary of the continuum extrapolations in one
dimension for mass ratio $m_{2}/m_{1}=1$. 

%TCIMACRO{\TeXButton{B}{\begin{table}[tbp] \centering}}%
%BeginExpansion
\begin{table}[H] \centering
%EndExpansion
\caption{Summary of continuum extrapolations for $m_{2}/m_{1}=1$ in one dimension}\label{1D_Extrap_ratio_1}%

$%
\begin{tabular}
[c]{|c|c|c|c|}%
$L_{z}|c|$ & $L_{z}$ & $R\kappa_{B}$ & $E_{K}/|E_{B}|
$\\\hline\hline

% Mass ratio 1 in 1D

20 & $200, 300, 400$ & $0.498$ & $0.006832$\\
10 & $100, 200, 300, 400$ & $0.4950$ & $0.030$\\
8 & $80, 160, 240$ & $0.4981$ & $0.050122$\\ 
10 & $100, 200, 300$ & $0.4812$ & $0.12083$\\
10 & $100, 200, 300$ & $0.461$ & $0.2694$

\end{tabular}
\ \ \ \ \ \ \ $\end{table}

Table \ref{1D_Extrap_ratio_2} gives the values of $L_{z}|c|$ and $L_{z}$ considered
as well as the continuum limit extrapolations in one dimension for mass ratio 
$m_{2}/m_{1}=2$.

%TCIMACRO{\TeXButton{B}{\begin{table}[tbp] \centering}}%
%BeginExpansion
\begin{table}[H] \centering
%EndExpansion
\caption{Summary of continuum extrapolations for $m_{2}/m_{1}=2$ in one dimension}\label{1D_Extrap_ratio_2}%

$%
\begin{tabular}
[c]{|c|c|c|c|}%
$L_{z}|c|$ & $L_{z}$ & $R\kappa_{B}$ & $E_{K}/|E_{B}|
$\\\hline\hline

% Mass ratio 2 in 1D

20 & $200, 300, 400$ & $0.6041$ & $0.0062$\\
10 & $100, 200, 300$ & $0.5957$ & $0.028228$\\
8 & $80, 160, 240$ & $0.5885$ & $0.0471$\\ 
10 & $100, 200, 300$ & $0.5676$ & $0.111164$\\
10 & $100, 200, 300$ & $0.5311$ & $0.247$

\end{tabular}
\ \ \ \ \ \ \ $\end{table}

In one dimension for mass ratio $m_{2}/m_{1}=4$ the continuum extrapolations
are given by Table \ref{1D_Extrap_ratio_4}.

%TCIMACRO{\TeXButton{B}{\begin{table}[tbp] \centering}}%
%BeginExpansion
\begin{table}[H] \centering
%EndExpansion
\caption{Summary of continuum extrapolations for $m_{2}/m_{1}=4$ in one dimension}\label{1D_Extrap_ratio_4}%

$%
\begin{tabular}
[c]{|c|c|c|c|}%
$L_{z}|c|$ & $L_{z}$ & $R\kappa_{B}$ & $E_{K}/|E_{B}|
$\\\hline\hline

% Mass ratio 4 in 1D

20 & $200, 300, 400$ & $0.8694$ & $0.00473$\\
10 & $100, 200, 300$ & $0.8504$ & $0.02292$\\
8 & $80, 160, 240$ & $0.8343$ & $0.039$\\ 
10 & $100, 200, 300$ & $0.7915$ & $0.0891$\\
10 & $100, 200, 300$ & $0.7214$ & $0.19409$

\end{tabular}
\ \ \ \ \ \ \ $\end{table}

In one dimension for mass ratio $m_{2}/m_{1}=8$ we get the continuum
extrapolations as shown in Table \ref{1D_Extrap_ratio_8}.

%TCIMACRO{\TeXButton{B}{\begin{table}[tbp] \centering}}%
%BeginExpansion
\begin{table}[H] \centering
%EndExpansion
\caption{Summary of continuum extrapolations for $m_{2}/m_{1}=8$ in one dimension}\label{1D_Extrap_ratio_8}%

$%
\begin{tabular}
[c]{|c|c|c|c|}%
$L_{z}|c|$ & $L_{z}$ & $R\kappa_{B}$ & $E_{K}/|E_{B}|
$\\\hline\hline

% Mass ratio 8 in 1D

20 & $200, 300, 400$ & $1.206$ & $0.003151$\\
10 & $100, 200, 300$ & $1.1742$ & $0.01665$\\ 
8 & $80, 160, 240$ & $1.1455$ & $0.029907$\\
10 & $100, 200, 300$ & $1.083$ & $0.0635$\\
10 & $100, 200, 300$ & $0.982$ & $0.13589$

\end{tabular}
\ \ \ \ \ \ \ $\end{table}
In two dimensions for mass ratios $m_{2}/m_{1}=1,2,4,8$ we consider the ground and 
first excited state for each $L_{z}\kappa_{B}=6.8,10.2$. As noted previously $L_{z}$ is the
length of the lattice in the direction perpendicular to the confining wall.
Table \ref{2D_Extrap_ratio_1} shows a summary of the continuum extrapolations in two
dimensions for mass ratio $m_{2}/m_{1}=1$. 

%TCIMACRO{\TeXButton{B}{\begin{table}[tbp] \centering}}%
%BeginExpansion
\begin{table}[h] \centering
%EndExpansion
\caption{Summary of continuum extrapolations for $m_{2}/m_{1}=1$ in two dimensions}\label{2D_Extrap_ratio_1}%

$%
\begin{tabular}
[c]{|c|c|c|c|}%
$L_{z}\kappa_{B}$ & $L_{z}$ & $R\kappa_{B}$ & $E_{K}/|E_{B}|
$\\\hline\hline

% Mass ratio 1 in 2D

10.2 & $60, 90, 120, 150$ & $0.485146$ & $0.004742$\\
6.8 & $40, 60, 80, 100$ & $0.456299$ & $0.011126$\\ 
10.2 & $60, 90, 120, 150$ & $0.482144$ & $0.019341$\\
6.8 & $40, 60, 80, 100$ & $0.45136$ & $0.044811$

\end{tabular}
\ \ \ \ \ \ \ $\end{table}

Table \ref{2D_Extrap_ratio_2} gives the values of $L_{z}\kappa_{B}$ and $L_{z}$ considered
as well as the continuum limit extrapolations in two dimensions for mass ratio 
$m_{2}/m_{1}=2$.

%TCIMACRO{\TeXButton{B}{\begin{table}[tbp] \centering}}%
%BeginExpansion
\begin{table}[h] \centering
%EndExpansion
\caption{Summary of continuum extrapolations for $m_{2}/m_{1}=2$ in two dimensions}\label{2D_Extrap_ratio_2}%

$%
\begin{tabular}
[c]{|c|c|c|c|}%
$L_{z}\kappa_{B}$ & $L_{z}$ & $R\kappa_{B}$ & $E_{K}/|E_{B}|
$\\\hline\hline

% Mass ratio 2 in 2D

10.2 & $60, 90, 120, 150$ & $0.572157$ & $0.004342$\\
6.8 & $40, 60, 80, 100$ & $0.541738$ & $0.010386$\\ 
10.2 & $60, 90, 120, 150$ & $0.567111$ & $0.017369$\\
6.8 & $40, 60, 80, 100$ & $0.532998$ & $0.040117$

\end{tabular}
\ \ \ \ \ \ \ $\end{table}

In two dimensions for mass ratio $m_{2}/m_{1}=4$ the continuum extrapolations
are given by Table \ref{2D_Extrap_ratio_4}.

%TCIMACRO{\TeXButton{B}{\begin{table}[tbp] \centering}}%
%BeginExpansion
\begin{table}[h] \centering
%EndExpansion
\caption{Summary of continuum extrapolations for $m_{2}/m_{1}=4$ in two dimensions}\label{2D_Extrap_ratio_4}%

$%
\begin{tabular}
[c]{|c|c|c|c|}%
$L_{z}\kappa_{B}$ & $L_{z}$ & $R\kappa_{B}$ & $E_{K}/|E_{B}|
$\\\hline\hline

% Mass ratio 4 in 2D

10.2 & $60, 90, 120, 150$ & $0.791197$ & $0.003473$\\
6.8 & $40, 60, 80, 100$ & $0.759959$ & $0.007583$\\ 
10.2 & $60, 90, 120, 150$ & $0.783424$ & $0.013027$\\
6.8 & $40, 60, 80, 100$ & $0.742787$ & $0.031096$

\end{tabular}
\ \ \ \ \ \ \ $\end{table}

In two dimensions for mass ratio $m_{2}/m_{1}=8$ we get the continuum
extrapolations as shown in Table \ref{2D_Extrap_ratio_8}.

%TCIMACRO{\TeXButton{B}{\begin{table}[tbp] \centering}}%
%BeginExpansion
\begin{table}[h] \centering
%EndExpansion
\caption{Summary of continuum extrapolations for $m_{2}/m_{1}=8$ in two dimensions}\label{2D_Extrap_ratio_8}%

$%
\begin{tabular}
[c]{|c|c|c|c|}%
$L_{z}\kappa_{B}$ & $L_{z}$ & $R\kappa_{B}$ & $E_{K}/|E_{B}|
$\\\hline\hline

% Mass ratio 8 in 2D
10.2 & $60, 90, 120, 150$ & $1.81456$ & $0.09496$\\
6.8 & $40, 60, 80, 100$ & $1.1018$ & $0.17597$\\ 
10.2 & $60, 90, 120, 150$ & $1.21421$ & $0.20091$\\
6.8 & $40, 60, 80, 100$ & $0.79017$ & $0.43994$

\end{tabular}
\ \ \ \ \ \ \ $\end{table}

In three dimensions we consider the ground and first excited states of $L_{z}\kappa_{B}=4.9,6$. 
Table \ref{3D_Extrap_ratio_1} shows a summary of the continuum extrapolations in three
dimensions for mass ratio $m_{2}/m_{1}=1$ 

%TCIMACRO{\TeXButton{B}{\begin{table}[tbp] \centering}}%
%BeginExpansion
\begin{table}[h] \centering
%EndExpansion
\caption{Summary of continuum extrapolations for $m_{2}/m_{1}=1$ in three dimensions}\label{3D_Extrap_ratio_1}%

$%
\begin{tabular}
[c]{|c|c|c|c|}%
$L_{z}\kappa_{B}$ & $L_{z}$ & $R\kappa_{B}$ & $E_{K}/|E_{B}|
$\\\hline\hline

% Mass ratio 1 in 3D

6 & $24, 36, 48$ & $0.27263$ & $0.08294$\\
4.9 & $20, 30, 40$ & $0.26891$ & $0.12394$\\
6 & $24, 36, 48$ & $0.25605$ & $0.32787$\\ 
4.94 & $20, 30, 40$ & $0.24773$ & $0.48669$

\end{tabular}
\ \ \ \ \ \ \ $\end{table}

Table \ref{3D_Extrap_ratio_2} gives the values of $L_{z}\kappa_{B}$ and $L_{z}$ considered
as well as the continuum limit extrapolations in three dimensions for mass ratio 
$m_{2}/m_{1}=2$.

%TCIMACRO{\TeXButton{B}{\begin{table}[tbp] \centering}}%
%BeginExpansion
\begin{table}[h] \centering
%EndExpansion
\caption{Summary of continuum extrapolations for $m_{2}/m_{1}=2$ in three dimensions}\label{3D_Extrap_ratio_2}%

$%
\begin{tabular}
[c]{|c|c|c|c|}%
$L_{z}\kappa_{B}$ & $L_{z}$ & $R\kappa_{B}$ & $E_{K}/|E_{B}|
$\\\hline\hline

% Mass ratio 2 in 3D

6 & $24, 36, 48$ & $0.32937$ & $0.0769$\\
4.9 & $20, 30, 40$ & $0.32308$ & $0.11573$\\
6 & $24, 36, 48$ & $0.30134$ & $0.30132$\\ 
4.9 & $20, 30, 40$ & $0.28801$ & $0.44853$

\end{tabular}
\ \ \ \ \ \ \ $\end{table}

In three dimensions for mass ratio $m_{2}/m_{1}=4$ the continuum extrapolations
are given by Table \ref{3D_Extrap_ratio_4}.

%TCIMACRO{\TeXButton{B}{\begin{table}[tbp] \centering}}%
%BeginExpansion
\begin{table}[h!] \centering
%EndExpansion
\caption{Summary of continuum extrapolations for $m_{2}/m_{1}=4$ in three dimensions}\label{3D_Extrap_ratio_4}%

$%
\begin{tabular}
[c]{|c|c|c|c|}%
$L_{z}\kappa_{B}$ & $L_{z}$ & $R\kappa_{B}$ & $E_{K}/|E_{B}|
$\\\hline\hline

% Mass ratio 4 in 3D

6 & $24, 36, 48$ & $0.47906$ & $0.06215$\\
4.9 & $20, 30, 40$ & $0.46579$ & $0.09545$\\
6 & $24, 36, 48$ & $0.42392$ & $0.23812$\\ 
4.9 & $20, 30, 40$ & $0.39833$ & $0.35779$

\end{tabular}
\ \ \ \ \ \ \ $\end{table}

In three dimensions for mass ratio $m_{2}/m_{1}=8$ we get the continuum
extrapolations as shown in Table \ref{3D_Extrap_ratio_8}.

%TCIMACRO{\TeXButton{B}{\begin{table}[tbp] \centering}}%
%BeginExpansion
\begin{table}[h] \centering
%EndExpansion
\caption{Summary of continuum extrapolations for $m_{2}/m_{1}=8$ in three dimensions}\label{3D_Extrap_ratio_8}%

$%
\begin{tabular}
[c]{|c|c|c|c|}%
$L_{z}\kappa_{B}$ & $L_{z}$ & $R\kappa_{B}$ & $E_{K}/|E_{B}|
$\\\hline\hline

% Mass ratio 8 in 3D
6 & $24, 36, 48$ & $0.68252$ & $0.0454$\\
4.9 & $20, 30, 40$ & $0.65848$ & $0.07191$\\
6 & $24, 36, 48$ & $0.59503$ & $0.16868$\\ 
4.9 & $20, 30, 40$ & $0.55407$ & $0.25769$

\end{tabular}
\ \ \ \ \ \ \ $\end{table}

\end{document}